\documentclass[twocolumn,twocolappendix]{aastex631}

\usepackage{amsmath,amstext}
\usepackage{bm}
\usepackage{mysymbol}
\usepackage{savesym}
\savesymbol{tablenum}
\usepackage{siunitx}
\restoresymbol{SIX}{tablenum}
\usepackage{threeparttable}
\usepackage{multirow}
\usepackage{scalefnt}
\usepackage{diagbox}

\accepted{January 15, 2023}

\submitjournal{ApJ}

\shorttitle{Probing Velocity Structures of Protostellar Envelopes}
\shortauthors{Sai et al.}
\graphicspath{{./}{figures/}}
\begin{document}

\title{Probing Velocity Structures of Protostellar Envelopes:\\ Infalling and Rotating Envelopes within Turbulent Dense Cores} %\footnote{Released on September, 1st, 2022}}

\correspondingauthor{Nagayoshi Ohashi}\email{ohashi@asiaa.sinica.edu.tw}

\author[0000-0003-4361-5577]{Jinshi Sai (Insa Choi)}
\affiliation{Department of Astronomy, Graduate School of Science, The University of Tokyo, 7-3-1 Hongo, Bunkyo-ku, Tokyo 113-0033, Japan}
\affiliation{Academia Sinica Institute of Astronomy and Astrophysics, 11F of Astro-Math Bldg, 1, Sec. 4, Roosevelt Rd, Taipei 10617, Taiwan}

\author[0000-0003-0998-5064]{Nagayoshi Ohashi}
\affiliation{Academia Sinica Institute of Astronomy and Astrophysics, 11F of Astro-Math Bldg, 1, Sec. 4, Roosevelt Rd, Taipei 10617, Taiwan}

\author[0000-0003-1412-893X]{Hsi-Wei Yen}
\affiliation{Academia Sinica Institute of Astronomy and Astrophysics, 11F of Astro-Math Bldg, 1, Sec. 4, Roosevelt Rd, Taipei 10617, Taiwan}

\author[0000-0002-3801-8754]{Ana\"{e}lle J. Maury}
\affiliation{AIM, CEA, CNRS, Universit\`{e} Paris-Saclay, Universit\`{e} Paris Diderot, Sorbonne Paris Cit\`{e}, 91191 Gif-sur-Yvette, France}
\affiliation{Harvard--Smithsonian Center for Astrophysics, Cambridge, MA02138, USA}

\author[0000-0003-1104-4554]{S\'{e}bastien Maret}
\affiliation{Univ. Grenoble Alpes, CNRS, IPAG, 38000 Grenoble, France}



\begin{abstract}

We have observed the three low-mass protostars, IRAS 15398$-$3359, L1527 IRS and TMC-1A, with the ALMA 12-m array, the ACA 7-m array, and the IRAM-30m and APEX telescopes in the C$^{18}$O $J=2$--1 emission. Overall, the C$^{18}$O emission shows clear velocity gradients at radii of $\sim$100--1000 au, which likely originate from rotation of envelopes, while velocity gradients are less clear and velocity structures are more perturbed on scales of $\sim$1000--10,000 au. IRAS 15398$-$3359 and L1527 IRS show a break at radii of $\sim$1200 and $\sim$1700 au in the radial profile of the peak velocity, respectively. The peak velocity is proportional to $r^{-1.38}$ or $r^{-1.7}$ within the break radius, which can be interpreted as indicating a rotational motion of the envelope with a degree of contamination of gas motions on larger spatial scales. The peak velocity follows $\vpeak \propto r^{0.68}$ or $\vpeak \propto r^{0.46}$ outside the break radius, which is similar to the $J/M$-$R$ relation of dense cores. TMC-1A exhibits the radial profile of the peak velocity not consistent with the rotational motion of the envelope nor the $J/M$-$R$ relation. The origin of the relation of $\vpeak \propto r^{0.46\operatorname{--}0.68}$ is investigated by examining correlations of the velocity deviation ($\delta v$) and the spatial scale ($\tau$) in the two sources. Obtained spatial correlations, $\delta v \propto \tau^{\sim0.6}$, are consistent with the scaling law predicted by turbulence models, which may suggest the large-scale velocity structures originate from turbulence.

\end{abstract}



\section{Introduction} \label{sec:intro}

Protoplanetary disks are sites of planet formation and are ubiquitously found around T Tauri stars \citep[e.g.,][]{Guilloteau:1998aa}. However, the detail of their formation process is still poorly understood. Circumstellar disks are expected to be formed around protostars as the angular momentum of dense cores is transferred to the center during the gravitational collapse of these cores \citep[][]{Terebey:1984aa}. Revealing the gas kinematics around protostars on scales ranging from dense cores to disks is thus essential to understand the physical processes of disk formation. 

The gas kinematics around protostars have been observationally investigated on various spatial scales. Earlier observational works on Class 0 and I protostars, especially with radio interferometers,
have revealed that infalling and rotational motions are dominant at scales of hundreds to thousands au \citep{Hayashi:1993aa, Momose:1998aa, Ohashi:1997aa, Hogerheijde:2001aa, DiFrancesco:2001aa, Belloche:2002aa, Arce:2004aa}. Recent observations at high angular resolutions suggest that radial distributions of rotational velocity at radii of $\sim$100--1000 au typically follow $v\propto r^{-1}$, i.e., the specific angular momentum $j$ is constant \citep[e.g,][]{Yen:2013aa, Harsono:2014aa, Ohashi:2014aa}, which is interpreted as a consequence of conservation of angular momentum during core collapse \citep{Takahashi:2016aa}. On the other hand, dense cores exhibit different velocity structures at scales of a few to tens of thousands au. \cite{Goodman:1993aa} have measured mean specific angular momentum $j=J/M$ of each dense core, where $J$ and $M$ is the total angular momentum and mass of the dense cores, respectively, from velocity gradients across sub-parsec scales assuming rigid-body rotation. They found that $J/M$ is proportional to $R^{1.6}$, where $R$ is the core size. This is known as the $J/M$-$R$ relation \citep[see also][]{Caselli:2002aa, Tatematsu:2016aa, Punanova:2018aa} Recently, \cite{Pineda:2019aa} revealed that the radial distributions of the specific angular momentum at radii of $\sim$800--10,000 au in protostellar and prestellar sources follow $j\propto r^{1.8}$, which is similar to the $J/M$-$R$ relation found in the earlier works.

Such observational works on different spatial scales have suggested that velocity structures at inner radii of $\sim$100--1000 au and outer radii of $\sim$1000-10,000 au of protostellar dense cores are distinct: the specific angular momentum $j$ is almost constant on the smaller scale but increases with increasing radius on the larger scale \citep[][]{Ohashi:1997ab, Belloche:2013aa}. \cite{Gaudel:2020aa} detected a transition between the two distinct regimes, i.e., the $j$-constant and $j$-increase regimes, at a radius of $\sim$1600 au for the first time by taking the average of radial distributions of the specific angular momentum of 12 Class 0 protostars. \cite{Sai:2022aa} reported a similar transition between the two regimes at a radius of $\sim$2900 au around the Class I protostar L1489 IRS, which is about two times larger than the averaged transitional radius reported by \cite{Gaudel:2020aa}.

Such a difference found in the two works raises a question of how different the transitional radius between the two regimes is from one source to another. Furthermore, the origin of the velocity structure at the $j$-increase regime is not well understood. Observations have shown complex velocity structures of dense cores on $\sim$1000-10,000 au scales : local velocity gradients within the dense cores show significant variation in direction. Such complex velocity structures cannot be interpreted as simple rigid-body rotation \citep{Caselli:2002aa, Tobin:2011aa, Chen:2019aa}. On one hand, simulation works suggest that the $J/M$-$R$ relation found in dense cores is inherited from turbulence in the parental filaments or clouds \citep[][]{Misugi:2019aa, Chen:2018aa}. The similarity between the $J/M$-$R$ relation ($v \sim j/r \propto r^{0.6}$) and the scaling law of the cloud-scale turbulence \citep[$\delta v\propto \tau^{0.5}$, where $\delta v$ is the velocity deviation and $\tau$ is the spatial scale; ][]{Larson:1981aa, McKee:2007aa} has also been pointed out \citep{Tatematsu:2016aa, Chen:2019ab, Pineda:2019aa, Gaudel:2020aa}. On the other hand, it is proposed that the $J/M$-$R$ relation arises as a consequence of gravitational contraction with angular momentum loss via turbulent viscosity \citep[][]{Arroyo-Chavez:2022aa}.

Turbulence possesses scale-dependent correlations of fluid variables on the spatial scale. This property of the turbulence is observed as a correlation between the velocity deviation $\delta v$ and the spatial scale $\tau$. A number of works have studied the correlation in molecular clouds with various methods and reported the scaling law of $\delta v\propto\tau^{\gamma}$, where $\gamma\sim$0.3--0.5, on $\sim$0.03--30 pc scales \citep[e.g.,][]{Larson:1981aa, Heyer:2004aa}. These indices are often interpreted as indicative of Kolmogorov turbulence, which yields $\gamma=1/3$ for incompressible fluids
\citep[][]{Kolmogorov:1941aa}, or Burgers turbulence, which yields $\gamma=1/2$ for highly compressible fluids \citep[][]{Burgers:1974aa}. However, such measurements have not been performed at smaller scales of a few to several thousands au inside dense cores because of the limit of angular and velocity resolutions.

In this paper, we present observations toward three well-studied protostars, i.e., two Class 0 protostars IRAS 15398--3359 and L1527 IRS (IRAS 04365$+$2557), and one Class I protostar TMC-1A (IRAS 04365$+$2535), with the ALMA 12-m array, the ACA 7-m array, and the IRAM-30m and Atacama Pathfinder Experiment (APEX) telescopes in the C$^{18}$O $J=2$--1 line. We investigate velocity structures around the protostars over a radius of $\sim$100--10,000 au with the C$^{18}$O maps. All the three protostars are located in the nearby star-forming regions: IRAS 15398$-$3359 is in the Lupus I molecular cloud \citep[$d\sim150$ pc;][]{Comeron:2008aa}, and L1527 IRS and TMC-1A are in the Taurus molecular cloud \citep[$d\sim140$ pc;][]{Zucker:2019aa}. Previous works have reported that rotational velocity of their envelopes is approximately proportional to $r^{-1}$ \citep{Yen:2013aa, Ohashi:2014aa, Aso:2015aa, Aso:2017ab, Yen:2017aa}. The Keplerian rotation curves of their disks were also identified from line observations, and dynamical masses were estimated \citep{Tobin:2012aa, Ohashi:2014aa, Aso:2017ab, Okoda:2018aa, Maret:2020aa}. The source properties are summarized in Table \ref{tab:summary_src}.

The outline of this paper is as follows. Observations and data reduction are summarized in Section \ref{sec:obs}. Observational results are presented in Section \ref{sec:results}. Analyses on the velocity structures are provided in Section \ref{sec:analysis}. The possible interpretation and implication of analysis on the velocity structures are discussed in Section \ref{sec:discussion}. Finally, all results and discussions are summarized in Section \ref{sec:summary}.

% ----------- Table: Source properties ---------
\begin{deluxetable*}{lccccccccc} %lccccccc
\tabletypesize{\footnotesize}
\centering
\tablecaption{Source properties \label{tab:summary_src}}
\tablehead{
\colhead{Source} & \colhead{R.A.} & \colhead{Decl.} & \colhead{$T_\mathrm{bol}$} & \colhead{$L_\mathrm{bol}$} & \colhead{$d$} & \colhead{$M_\ast$} & \colhead{$R_\mathrm{disk}$} & \colhead{$i$} & \colhead{References}  \\
\colhead{} & \colhead{(J2000)} & \colhead{(J2000)} & \colhead{(K)} & \colhead{($\Lsun$)} & \colhead{(pc)} & \colhead{($\Msun$)} & \colhead{(au)} & ($^\circ$) & \colhead{} }
%\decimalcolnumbers
\startdata
IRAS 15398$-$3359 & 15:43:02.24 & $-$34:09:06.81 & 61 & 0.92 & 150 & 0.007 & $\gtrsim$40 & 70 & 1, 2, 3, 4, 5 \\
L1527 IRS & 4:39:53.88 & $+$26:03:09.55 & 44 & 2.0 & 140 & 0.45 & 74 & 85--90 & 6, 7, 8, 9 \\
TMC-1A & 4:39:35.20 & $+$25:41:44.35 & 118 & 2.7 & 140 & 0.68 & 100 & 55--65 & 6, 7, 10, 11
\enddata
\tablecomments{(1) \cite{Froebrich:2005aa}; (2) \cite{Comeron:2008aa}; (3) \cite{Okoda:2018aa}; (4) \cite{Oya:2014aa}; (5) \cite{Yen:2017aa}; (6) \cite{Kristensen:2012aa}; (7) \cite{Zucker:2019aa}; (8) \cite{Aso:2017ab}; (9) \cite{Tobin:2008aa}; (10) \cite{Aso:2015aa}; (11) \cite{Harsono:2014aa}.}
\end{deluxetable*}
\vspace{-6mm}
% ----------------------------------------

\section{Observations} \label{sec:obs}
\subsection{ALMA 12-m array Observations \label{subsec:obs_alma}}
The three protostars were observed with the Atacama Large Millimeter/submillimeter Array (ALMA) 12-m array in the C$^{18}$O $J=2$--1 (219.560358 GHz; $E_\mathrm{up}=15.8$ K; $A_\mathrm{ul}=6.01\times 10^{-7}$ s$^{-1}$) line emission in the ALMA Cycle 2. The ALMA observations are summarized in Table \ref{tab:obssummary_alma}. The observations consist of a single field for each source, and the field of view is $\ang[angle-symbol-over-decimal]{;;26}$ at 220 GHz. The observations of L1527 IRS and TMC-1A were conducted with a compact configuration with 34 antennas and an extended configuration with 35 antennas. The projected baseline lengths of compact and extended configurations were 12--363 k$\lambda$ (16--494 m) and 21--1499 k$\lambda$ (29--2043 m) at 220 GHz, respectively. The spectral window for the C$^{18}$O $J=2$--1 line had a bandwidth of 58.6 MHz and a spectral resolution of 30.5 kHz, corresponding to a velocity resolution of 0.042 $\kmps$. The total on-source time was 48 min for L1527 IRS and 46 min for TMC-1A. The data obtained with compact and extended antenna configurations were calibrated with the Common Astronomy Software Applications package \citep[CASA;][]{McMullin:2007aa} version 4.5.0 and 4.5.2, respectively. IRAS 15398$-$3359 was observed with a single configuration with 34 antennas. The shortest and longest projected baseline length were 13 k$\lambda$ and 478 k$\lambda$ (18 m and 651 m) at 220 GHz, respectively. The spectral window for the C$^{18}$O $J=2$--1 line had a bandwidth of 58.6 MHz and a spectral resolution of 61.0 kHz ($\sim$0.084 $\kmps$). The on-source time was 88 min. The detail of the observations are summarized in \cite{Yen:2017aa}. The data was calibrated with CASA 4.3.1.

% ----------- Table: summary of ALMA observations ---------
\begin{deluxetable*}{lcccccc} %lccccccc
\tabletypesize{\scriptsize}
\centering
\tablecaption{Summary of ALMA observations \label{tab:obssummary_alma}}
\tablehead{
\colhead{Source} & \colhead{R.A.$^{1}$} & \colhead{Decl.$^{1}$} & \colhead{Baseline length$^{2}$} & Date & \colhead{Calibrators} & \colhead{$\Delta v$}  \\
\colhead{} & \colhead{(J2000)} & \colhead{(J2000)} & \colhead{(k$\lambda$)} & \colhead{(UT)} & \colhead{(Gain, Bandpass, Flux)} & \colhead{($\kmps$)} }
%\decimalcolnumbers
\startdata
IRAS 15398$-$3359 & 15:43:02.16 & $-$34:09:09.0 & 13--478 & Apr.~30, May 19, Jun.~6, 2014 & J1534$-$3526, J1427$-$4206, Titan & 0.084 \\
L1527 IRS & 04:39:53.91 & $+$25:41:44.4 & 12--363 & May 24, 2015 & J0510$+$1800, J0423$-$0120, J0510$+$180 & 0.042 \\
 & & & 21--1499 & Sep. 20, 2015 & J0429$+$2724, J0510$+$1800, J0423$-$013 & 0.042 \\
TMC-1A & 04:39:35.20 & $+$25:41:44.4 & 12--363 & May 24, 2015 & J0510$+$1800, J0423$-$0120, J0510$+$180 & 0.042 \\
 & & & 21--1499 & Sep. 20, 2015 & J0429$+$2724, J0510$+$1800, J0423$-$013 & 0.042
\enddata
\tablecomments{$^1$R.A. and Decl. of the phase center of the observations; $^2$Projected baseline lengths at 220 GHz.}
\end{deluxetable*}
\vspace{-6mm}
% ----------------------------------------

\subsection{ACA 7-m Array Mosaic Observations}
We have conducted mosaic observations of the three protostars in the C$^{18}$O $J=2$--1 emission using the 7-m array of the Atacama Compact Array (ACA) during December 1--19, 2019 in the ALMA Cycle 7 with 10 antennas. The summary of the observations are presented in Table \ref{tab:obssummary_aca7m}. The mosaic observations covered $\sim\ang[angle-symbol-over-decimal]{;2;}\times\ang[angle-symbol-over-decimal]{;2;}$ regions centered at protostellar positions. The spectral window for the C$^{18}$O $J=2$--1 line had a bandwidth of 120.0 MHz and a spectral resolution of 61.0 kHz, providing a velocity resolution of 0.084 $\kmps$. The mosaic maps of IRAS 15398, L1527 IRS and TMC-1A consist of 28, 27 and 25 fields and the on-source time on each single field was 1.7, 1.1 and 3.2 min, respectively. The projected baseline lengths ranged from 5 k$\lambda$ to 28 k$\lambda$ (6.8--38 m) for L1527 IRS and TMC-1A, and from 6 k$\lambda$ to 35 k$\lambda$ (8.2--48 m) for IRAS 15398--3359 at 220 GHz. The data was calibrated in the pipeline with CASA 5.6.1 and its pipeline version was 42866M (Pipeline-CASA56-P1-B).

% ----------- Table: summary of ACA 7m observations ---------
\begin{deluxetable*}{lccccc} %lccccccc
\tabletypesize{\footnotesize}
\centering
\tablecaption{Summary of ACA 7-m array observations \label{tab:obssummary_aca7m}}
\tablehead{
\colhead{Source} & \colhead{Number of fields} & \colhead{Date} & \colhead{Gain Calibrators} &
\colhead{Bandpass and flux calibrators$^{1}$} & \colhead{$\Delta v$}  \\
\colhead{} & \colhead{} & \colhead{(UT)} & \colhead{} &
\colhead{} & \colhead{($\kmps$)} }
%\decimalcolnumbers
\startdata
IRAS 15398$-$3359 & 28 & Dec.~19, 2019 & J1534$-$3526 & J1337$-$1257  & 0.084 \\
L1527 IRS & 27 & Dec.~1--17, 2019 & J0426$+$2327 & J0423$-$0120/J0725$-$0054 & 0.084 \\
TMC-1A & 25 & Dec.~1--17, 2019 & J0426$+$2327 & J0423$-$0120/J0725$-$0054 & 0.084
\enddata
\tablecomments{$^1$Bandpass and flux were calibrated using the same source, either of J1337$-$1257, J0423$-$0120 or J0725$-$0054 depending on the target source and observing date. }
\end{deluxetable*}
% ----------------------------------------

\subsection{Single-dish Observations}

% ----------- Table: summary of SD observations ---------
\begin{deluxetable*}{lccccccc} %lccccccc
\tabletypesize{\footnotesize}
\centering
\tablecaption{Summary of single-dish observations \label{tab:obssummary_sd}}
\tablehead{
\colhead{Source} & \colhead{Telescope} & \colhead{Receiver/Backend} & \colhead{Date} & \colhead{$\Delta v$} & \colhead{Mapping area} \\
\colhead{} & \colhead{} & \colhead{} & \colhead{(UT)} & \colhead{($\kmps$)} & }
%\decimalcolnumbers
\startdata
IRAS 15398$-$3359 & APEX & APEX-1/XFFTS & Jul.~31--Aug.~2, 2017 & 0.01 & $4.3'\times 4.3'$ \\
L1527 IRS & IRAM-30m & HERA/VESPA & Sep.~3--4, 2014 & 0.027 & $2.2'\times 2.2'$ \\
TMC-1A & IRAM-30m & EMIR/VESPA & Dec.~27--31, 2019 & 0.027 & $2.7'\times 2.5'$
\enddata
%\tablecomments{$^1$}
\end{deluxetable*}
\vspace{-6mm}
% ----------------------------------------

Observations of the three protostars with the IRAM-30m and APEX telescopes have been carried out in the C$^{18}$O $J=2$--1 emission in the on-the-fly (OTF) mapping mode to cover $2'$ square regions. The position-switching method was used for the OTF mapping in all observations. The summary of the observations is presented in Table \ref{tab:obssummary_sd}.

L1527 IRS was observed with the IRAM-30m telescope using the Heterodyne Receiver Array (HERA) receiver and the Versatile Spectrometer Arrays (VESPA) backend with a spectral resolution of 20 kHz (0.027 $\kmps$). The pointing was corrected every two hours with quasars near the target: 0316+413 and 0415+379. The focus was corrected every four hours with the quasar 0316+413. The reference position for the position-switching was $\Delta\alpha=6'$, $\Delta\delta=-12'$ with respect to the source position given in Table \ref{tab:summary_src}.

TMC-1A was observed with the IRAM-30m telescope using the heterodyne receiver Eight MIxer Receiver (EMIR) E230 with the VESPA backend with a spectral resolution of 20 kHz (0.027 $\kmps$). The telescope pointing and focus were corrected every $\sim$2--3 and $\sim$4--5 hours with the quasars 0430+052 and 0316+413, respectively. The reference position for the position-switching was $\Delta\alpha=32'$, $\Delta\delta=48'$ with respect to the source position.

Observations of IRAS 15398$-$3359 was conducted using the APEX telescope on July 31 to August 2, 2017. The APEX-1 heterodyne receiver was used together with the RPG extended bandwidth fast Fourier transform spectrometer (XFFTS) backend. The spectral resolution was 76 kHz (0.1 $\kmps$). The pointing was corrected every $\sim$1--2 hours with RAFGL 4211 and IRAS 15194$-$5115. The reference position for the position-switching was $\Delta\alpha=6'$, $\Delta\delta=19'$ with respect to the protostellar position.

All single-dish data were reduced with the GILDAS software package. The antenna temperature $\ta$ was converted to the main beam temperature $\tmb$ according to the relation $\tmb = \ta \feff/\beff$, where $\beff$ is the main beam efficiency and $\feff$ is the forward efficiency. We adopted $\beff=0.60$ and $\feff=0.92$ for the C$^{18}$O $J=2$--1 data obtained with IRAM-30m, and $\beff=0.63$ and $\feff=0.95$ for the APEX data. The half-power beam width (HPBW) at the rest frequency of C$^{18}$O $J=2$--1 is $\sim\ang[angle-symbol-over-decimal]{;;12}$ for the IRAM-30m maps and $\sim\ang[angle-symbol-over-decimal]{;;30}$ for the APEX map.

\subsection{Combined Maps \label{subsec:obs_combmaps}}
The interferometric and single-dish data were combined for further analysis with the task \texttt{feather} in CASA 5.6.1, which combines interferometric and single-dish images in Fourier space, weighting them by the spatial frequency response of each image. Two maps with different map sizes and angular resolutions were generated for each source to trace different spatial scales: (i) combined maps of the 12-m array, 7-m array and single-dish data at high angular resolutions to resolve structures near protostars (within a radius of $\sim\ang[angle-symbol-over-decimal]{;;12}$) and (ii) combined maps of the 7-m array and single-dish data covering wider areas around protostars ($\sim\ang[angle-symbol-over-decimal]{;2;} \times \ang[angle-symbol-over-decimal]{;2;}$). We compared fluxes of the data in both the image and visibility domain, and confirmed that all data are consistent in flux. To compare the single-dish data in the visibility domain, pseudo-visibilities were generated using the MIRIAD software following the methods described in \cite{Takakuwa:2003aa}. The peak positions of continuum emission observed with ALMA in previous works were adopted as protostellar positions and map centers \citep{Aso:2015aa, Aso:2017ab, Yen:2017aa}. The protostellar positions are summarized in Table \ref{tab:summary_src}. All the imaging process were performed using CASA 5.6.1. The angular resolutions, rms, velocity resolutions of the produced maps are summarized in Table \ref{tab:mapsummary}.

% ----------- Table: summary of maps ---------
\begin{table*}[thb]
\centering
\caption{Summary of C$^{18}$O maps}
\begin{tabular*}{2\columnwidth}{@{\extracolsep{\fill}}llccc}
\hline
\hline
Map & & IRAS 15398--3359 & L1527 IRS & TMC-1A \\
\hline
Small scale & Angular resolution & \(\ang[angle-symbol-over-decimal]{;;0.77}\times\ang[angle-symbol-over-decimal]{;;0.72}~(75^\circ)\) &
\(\ang[angle-symbol-over-decimal]{;;0.94}\times \ang[angle-symbol-over-decimal]{;;0.82}~(8.1^\circ)\) & \(\ang[angle-symbol-over-decimal]{;;0.93}\times \ang[angle-symbol-over-decimal]{;;0.81}~(7.6^\circ)\) \\
& rms (\(\mjypbm\)) & 5.2 & 12 & 12 \\
& Velocity resolution (\(\kmps\)) & 0.11 & 0.084 & 0.084 \\
\hline
Large scale & Angular resolution & \(\ang[angle-symbol-over-decimal]{;;8.1}\times\ang[angle-symbol-over-decimal]{;;4.4}~(85^\circ)\) &
\(\ang[angle-symbol-over-decimal]{;;7.5}\times\ang[angle-symbol-over-decimal]{;;6.4}~(-1.5^\circ)\) & \(\ang[angle-symbol-over-decimal]{;;7.7}\times\ang[angle-symbol-over-decimal]{;;6.4}~(-85^\circ)\) \\
& rms (\(\mjypbm\)) & 160 & 290 & 300 \\
& Velocity resolution (\(\kmps\)) & 0.11 & 0.084 & 0.084 \\
\hline
\label{tab:mapsummary}
\end{tabular*}
\end{table*}
% ----------------------------------------

\subsubsection{Small-scale Maps}
Maps covering regions near protostars at high angular resolutions were produced by combining the 12-m array, 7-m array and single-dish data. We have referred to these maps as small-scale maps throughout this paper. First, the combined maps of 12-m and 7-m array data were produced through joint deconvolutions with the task \texttt{tclean}. To enhance the sensitivity to extended structures, the natural weighting and taper with the FWHM of 200 k$\lambda$ were adopted. Multiscale CLEAN was applied with scale sizes of zero, $\ang[angle-symbol-over-decimal]{;;1}$ and $\ang[angle-symbol-over-decimal]{;;3}$. Then, the 12-m+7-m maps were combined with the single-dish maps using the following steps: (1) the single-dish and 12-m+7-m images were trimmed to exclude masked regions at the map edges; (2) the single-dish images were multiplied by the 12-m+7-m primary beam response; (3) the modified single-dish maps were combined with the 12-m+7-m maps with the task \texttt{feather}; and (4) the combined maps were corrected by the 12-m+7-m primary beam response. As such, the angular resolution, rms noise level and velocity resolution are $\sim\ang[angle-symbol-over-decimal]{;;0.8}$, $\sim$11 $\mjypbm$ and $0.09~\kmps$, respectively, on average. The final map size centered at the protostellar positions is $\sim$24$''$ square.

\subsubsection{Large-scale Maps}
Combined maps of the 7-m array and single-dish data were produced to cover wider regions. We have referred to these maps as large-scale maps throughout this paper. First, mosaic maps of the 7-m array data were produced using the \texttt{tclean} task with a Briggs robust parameter of 0.5 and the Hogbom deconvolver. The cleaned 7-m maps were combined with the single-dish maps using the \texttt{feather} task and following the same steps used to produce the small-scale maps. The angular resolution, rms noise level and velocity resolution of the final maps are $\sim\ang[angle-symbol-over-decimal]{;;7}$, $\sim$25 $\mjypbm$ and $0.09~\kmps$, respectively, on average. The maps cover $\sim\ang[angle-symbol-over-decimal]{;2;}$ square regions centered at the protostars.

\section{Results} \label{sec:results}

% Overall
The C$^{18}$O $J=2$--1 integrated intensity and centroid velocity maps of the three protostars are presented in Figure \ref{fig:momentmaps}, where small-scale and large-scale maps are shown in the left and the right columns, respectively. For the integrated intensity maps, the C$^{18}$O emissions were integrated over velocity ranges where the emission was detected at least above 3$\sigma$. For the centroid velocity maps, only emission detected above 5$\sigma$ was considered. In both small- and large-scale maps, the emission peaks are located at the protostellar positions. Thus, the C$^{18}$O emission traces the envelopes associated with the protostars. Figures \ref{fig:channel_iras15398} to \ref{fig:channel_ls_tmc1a} in the Appendix show velocity channel maps for each protostar, and more details of moment 0 maps are presented below.

% IRAS 15398
\subsection{IRAS 15398--3359\label{subsec:res_iras15398}}
In the small-scale map in Figure \ref{fig:momentmaps}(a), the distribution of the C$^{18}$O integrated intensity exhibits an X-shape morphology: a part of the emission is elongated from northeast to southwest, while another part is along a direction from southeast to northwest. These elongated structures are $\sim\ang[angle-symbol-over-decimal]{;;15}$ in length across the protostellar position. The elongation from the northeast to southwest is the same as the direction of the primary outflow associated with the protostar \citep{Oya:2014aa, Yen:2017aa}, and it exhibits a velocity gradient along the elongation. The velocity gradient showing redshifted velocity on its northeastern side and blueshifted velocity on its southwestern side is also the same as that of the primary outflow. These suggest that the emission elongated from the northeast to southwest is likely affected by the primary outflow. Weak emission extends to southeast from the protostar, which is almost the same direction as that of the secondary outflow suggested in a previous work \citep{Okoda:2021aa}. Outside the X-shape morphology, the C$^{18}$O emission extends over the entire map without any clear velocity gradient. In the large-scale map shown in Figure \ref{fig:momentmaps}(b), a weak velocity gradient is seen from northwest to southeast, although the velocity structure appears more perturbed than in the small-scale map.

While the systemic velocity of this system has been estimated based on rotational velocity of the envelope and disk at radii less than 100 au \citep[][]{Yen:2017aa, Okoda:2018aa}, the values reported in the previous works may not be appropriate at the spatial scales of the current maps because the center of mass could be different due to the very small protostellar mass of this source \citep[$\sim$0.007 $\Msun$;][]{Okoda:2018aa}. Indeed, \cite{Okoda:2018aa} reported a discrepancy between systemic velocities estimated from rotational velocity of the disk and the envelope. Thus, we estimate the systemic velocity to be 5.183$\pm0.005$ $\kmps$ using our large-scale map by fitting a Gaussian function to the C$^{18}$O spectrum at the protostellar position. We have adopted this centroid velocity of 5.18 $\kmps$ for the systemic velocity of this system in the following analyses and discussions.

\subsection{L1527 IRS\label{subsec:res_l1527}}
\begin{figure*}[htbp]
\centering
\includegraphics[width=2\columnwidth]{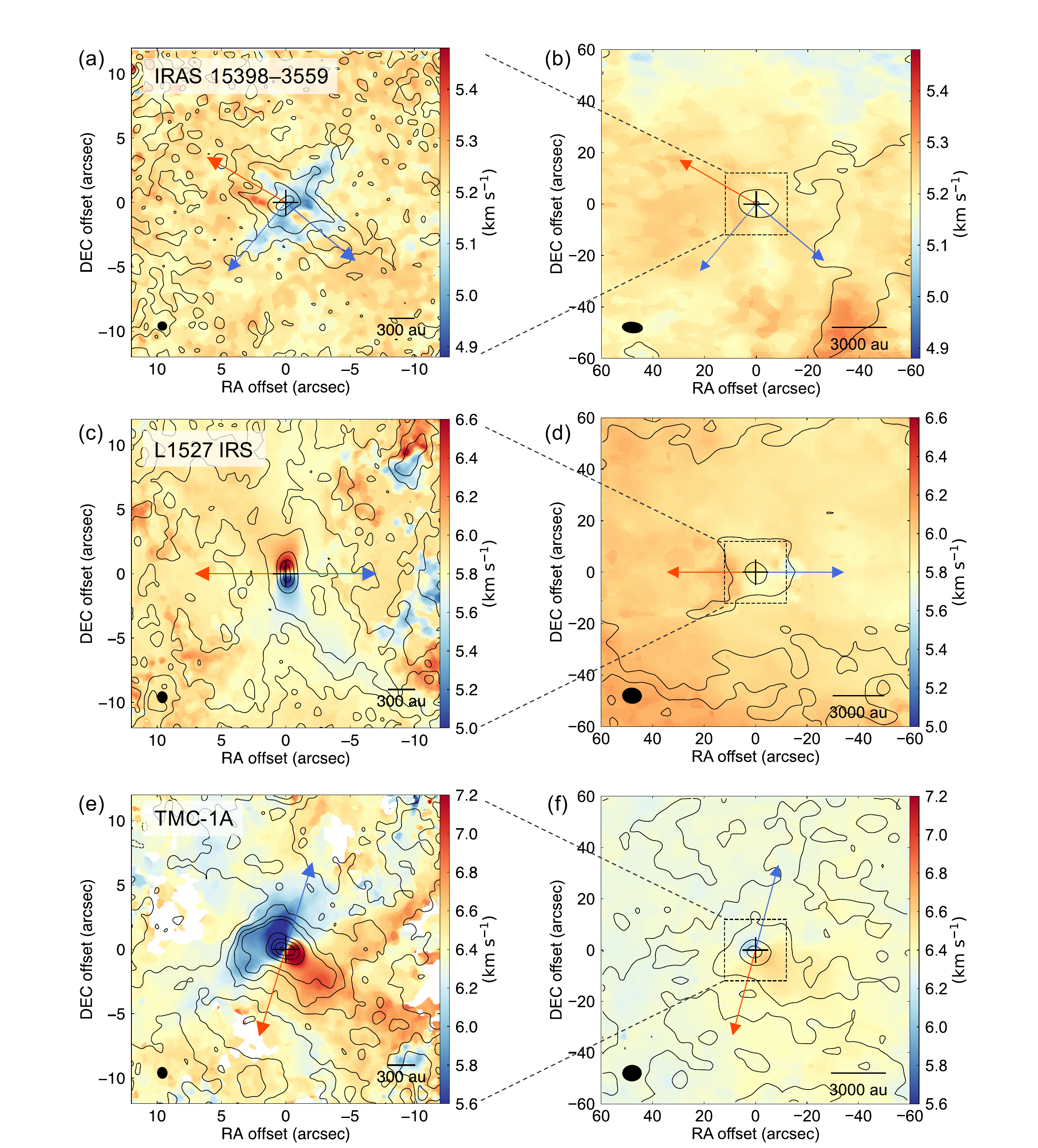}
\caption{Integrated intensity (contour) and centroid velocity (color) maps of the C$^{18}$O $J=2$--1 emissions of IRAS 15398$-$3359, L1527 IRS and TMC-1A. The left and right columns show small- and large-scale maps, respectively. In the maps of the left column, contour levels are from 5$\sigma$ to 20$\sigma$ in steps of 5$\sigma$, and then from 20$\sigma$ to 80$\sigma$ in steps of 20$\sigma$, where 1$\sigma$ is the rms noise of a map for a given source, which is summarized in Table \ref{tab:mapsummary}. In the maps of the right column, the contour steps are the same as those in the maps of the left column, but begin from 40$\sigma$ in the map of IRAS 15398$-$3359 and from 10$\sigma$ in the maps of L1527 IRS and TMC-1A. Red and blue arrows with solid lines show directions of (primary) outflows. Blue arrows with dashed lines indicate the direction of the secondary outflow of IRAS 15398$-$3359. Black crosses at the center and ellipses in the bottom-left corners denote the protostellar positions and the synthesized beam size, respectively.}
\label{fig:momentmaps}
\end{figure*}

Figure \ref{fig:momentmaps}(c) and (d) shows the integrated intensity and centroid velocity maps of L1527 IRS. The small-scale map in Figure \ref{fig:momentmaps}(c) shows a flattened structure along the north-south direction within a radius of $\sim\ang[angle-symbol-over-decimal]{;;2}$ around the protostar. A clear velocity gradient is seen along the flattened structure. The velocity gradient consists of velocity components blueshifted and redshifted with respect to the systemic velocity of the protostar \citep[$5.8~\kmps$;][]{Aso:2017ab} to the south and north sides of the protostar, respectively. This feature suggests that the emission likely traces the rotational motion of the disk and envelope, as found in previous works \citep{Tobin:2012aa, Sakai:2014ab, Ohashi:2014aa, Aso:2017ab}. Weak emission is elongated from northeast to southwest over $\sim\ang[angle-symbol-over-decimal]{;;20}$ across the protostellar position outside the flattened emission. Another component is also extended from the protostar toward the southeast. These structures seem to be associated with its outflow cavity, as the outflow has been detected along the east--west direction \citep{Tamura:1996aa,Tobin:2008aa}. In the large-scale map in Figure \ref{fig:momentmaps}(d), the emission is elongated and exhibits a weak velocity gradient in the east--west direction, which likely traces the outflow (see also velocity channel maps in Figure \ref{fig:channel_ls_l1527}). On the other hand, the velocity gradient in the direction perpendicular to the outflow becomes less clear in the large-scale map.

% -------- Figures: results --------------
% Moment maps
%\begin{landscape}
%\end{landscape}
% ----------------------------------

\subsection{TMC-1A\label{sec:res_tmc1a}}
In the small-scale map of TMC-1A in Figure \ref{fig:momentmaps}(e), the C$^{18}$O emission near the protostar has a flattened shape and a clear velocity gradient in a direction from northeast to southwest, which is almost perpendicular to the outflow. The velocity gradient consists of blueshifted and redshifted velocities with respect to the systemic velocity \citep[$6.4~\kmps$;][]{Aso:2015aa} on the northeast and southwest sides of the protostar, respectively. These features are consistent with previous ALMA observations, which suggested that these velocity structures originate from the rotating disk and envelope \citep{Aso:2015aa}. Outside the flattened emission, weak emission is extended from northeast to southwest and from southeast to northwest. These components are considered to trace its outflow cavity, as the outflow was detected along the southeast to the northwest \citep{Aso:2015aa}. A weak velocity gradient from northeast to southwest is observed in the large-scale map in Figure \ref{fig:momentmaps}(f), although it appears more perturbed.
%

% --------- Analysis ---------
\section{Analysis\label{sec:analysis}}

% two-D velocity field fitting
\subsection{Two Dimensional Velocity Gradient \label{subsec:ana_vgrad}}
To characterize the observed velocity gradients quantitatively, the following two-dimensional linear function, introduced by \cite{Goodman:1993aa}, was fitted to the centroid velocity maps with different aperture areas representing different spatial scales:
\begin{align}
    \vlsr = v_0 + a \Delta \alpha + b \Delta \delta,
\end{align}
where $\Delta \alpha$ and $\Delta \delta$ are offsets from the protostellar position in right ascension $\alpha$ and declination $\delta$, respectively, and $v_0$, $a$ and $b$ are the fitting parameters. The magnitude and position angle of the velocity gradient are calculated as follows:
\begin{align}
    G & = \left(a^2  + b^2 \right)^{1/2}/d, \label{eq:mag_grad} \\
    \theta & = \tan^{-1}(a/b), \label{eq:pa_grad}
\end{align}
where $G$ is the magnitude, $\theta$ is the position angle measured from north to east, and $d$ is the distance of the source. The small-scale maps were used for the fitting with the aperture radius $r_\mathrm{fit}$ of $\ang[angle-symbol-over-decimal]{;;5}$--$\ang[angle-symbol-over-decimal]{;;10}$, while the fitting with $r_\mathrm{fit}$ of $\ang[angle-symbol-over-decimal]{;;20}$--$\ang[angle-symbol-over-decimal]{;;60}$ was performed using the large-scale maps.

The fitting results are summarized in Table \ref{tab:vgrad}. IRAS 15398$-$3359 exhibits the weakest velocity gradient among the three protostars on all scales. TMC-1A shows the strongest velocity gradient near the protostar within a radius of $\ang[angle-symbol-over-decimal]{;;30}$, while L1527 IRS does on scales larger than $\ang[angle-symbol-over-decimal]{;;30}$. Overall, the velocity gradients are stronger near the protostars and weaker at larger radii in all sources. The same fitting has been performed in L1527 IRS by \cite{Maret:2020aa} and \cite{Gaudel:2020aa} using different data sets. \cite{Gaudel:2020aa} report $G=66~\kmps~\mathrm{pc}^{-1}$ and $\theta=22^\circ$ within a radius of $5''$ using a C$^{18}$O $J=2$--1 map of PdBI$+$IRAM-30m data, which is slightly different from the values measured in the current work. These differences arise because our map has better sensitivity to extended emission, and more pixels with low velocities close to the systemic velocity are included even in the fitting within a radius of $5''$. On the other hand, the centroid velocity map presented by \cite{Gaudel:2020aa} has blank pixels at the map edge because of the sensitivity limit, excluding extended, low-velocity components from the fitting. They measure $G=2~\kmps~\mathrm{pc}^{-1}$ and $\theta=113^\circ$ within a radius of $40''$ using a IRAM-30m C$^{18}$O $J=2$--1 map, which is in agreement with our results. \cite{Maret:2020aa} present $G=671~\kmps~\mathrm{pc}^{-1}$ within a radius of $2''$, which also agrees with the trend that the velocity gradient is stronger at smaller radii.

Scale dependence of the directions of the velocity gradients differs from source to source. The position angles of the velocity gradients of IRAS 15398$-$3359 within a radius of $\ang[angle-symbol-over-decimal]{;;10}$ is $\sim$70$^\circ$, which is a direction of the associated outflow \citep{Oya:2014aa, Yen:2017aa}. The directions of the velocity gradients of IRAS 15398$-$3359 vary significantly from 70$^\circ$ to 170$^\circ$ as the scale becomes larger. In L1527 IRS, the position angle of the velocity gradient within a radius of $\ang[angle-symbol-over-decimal]{;;5}$ is $\sim$14$^\circ$, which is close to the direction of the disk major axis \citep[$\mathrm{P.A.}=0^\circ$;][]{Tobin:2012ab, Ohashi:2014aa, Aso:2017ab}. L1527 IRS also shows significant change of the directions of the velocity gradients by $\sim$90$^\circ$ as the scale becomes larger. The velocity gradient over a radius of $\ang[angle-symbol-over-decimal]{;;60}$ is along P.A.$\sim$110$^\circ$, which is close to the outflow direction \citep[$\mathrm{P.A.}=90^\circ$; ][]{Hogerheijde:1998aa, Tobin:2008aa, Oya:2015aa}. On the contrary, TMC-1A shows a small dispersion of the directions of the velocity gradients $\sim\pm10^\circ$ over spatial scales of $\ang[angle-symbol-over-decimal]{;;5}$--$\ang[angle-symbol-over-decimal]{;;60}$. The directions of the velocity gradients are roughly normal to the outflow with $\mathrm{P.A.}=-25$--$-17^\circ$ \citep{Hogerheijde:1998aa, Aso:2015aa}. % within $\sim$10--30$^\circ$.

% ----------- Table: results ------------
% Velocity gradients
%\begin{longrotatetable}
\begin{table*}[thbp]
\begin{threeparttable}
    \centering
    \caption{Results of the two-dimensional linear fitting to the centroid velocity maps}
    \label{tab:vgrad}
    \scalefont{0.82}
    \begin{tabular*}{2\columnwidth}{@{\extracolsep{\fill}}lccccccccc}
    \hline
    \hline
     & \multicolumn{3}{c}{IRAS 15398$-$3359} & \multicolumn{3}{c}{L1527 IRS} & \multicolumn{3}{c}{TMC-1A} \\
    $r_\mathrm{fit}$ & $v_0$\tnote{a} & $G$ & $\theta$ & $v_0$ & $G$ & $\theta$ & $v_0$ & $G$ & $\theta$ \\
     ($''$) & ($\mathrm{km~s^{-1}}$) & ($\mathrm{km~s^{-1}~pc^{-1}}$) & ($^\circ$) & ($\mathrm{km~s^{-1}}$) & ($\mathrm{km~s^{-1}~pc^{-1}}$) & ($^\circ$) & ($\mathrm{km~s^{-1}}$) & ($\mathrm{km~s^{-1}~pc^{-1}}$) & ($^\circ$)\\
    \hline
    5 & 5.18 & 14.1 $\pm$ 0.2 & 69.3 $\pm$ 0.8 & 5.83 & 41.7 $\pm$ 0.2 & 14.0 $\pm$ 0.3 & 6.34 & 107.8 $\pm$ 0.2 & -146.18 $\pm$ 0.09 \\
    10 & 5.21 & 1.05 $\pm$ 0.05 & 67 $\pm$ 3 & 5.87 & 8.39 $\pm$ 0.04 & 5.8 $\pm$ 0.3 & 6.39 & 18.50 $\pm$ 0.03 & -125.9 $\pm$ 0.1 \\
    20 & 5.23 & 1.6 $\pm$ 0.1 & 83 $\pm$ 4 & 5.91 & 3.7 $\pm$ 0.1 & 96 $\pm$ 2 & 6.40 & 8.78 $\pm$ 0.08 & -149.6 $\pm$ 0.5 \\
    30 & 5.22 & 0.97 $\pm$ 0.05 & 96 $\pm$ 3 & 5.93 & 2.26 $\pm$ 0.04 & 108 $\pm$ 1 & 6.38 & 4.17 $\pm$ 0.03 & -144.3 $\pm$ 0.4 \\
    40 & 5.22 & 0.89 $\pm$ 0.03 & 107 $\pm$ 2 & 5.94 & 2.66 $\pm$ 0.02 & 112.9 $\pm$ 0.5 & 6.37 & 2.49 $\pm$ 0.02 & -135.5 $\pm$ 0.4 \\
    50 & 5.22 & 0.71 $\pm$ 0.02 & 137 $\pm$ 1 & 5.94 & 2.56 $\pm$ 0.01 & 113.7 $\pm$ 0.3 & 6.36 & 1.72 $\pm$ 0.01 & -129.2 $\pm$ 0.4 \\
    60 & 5.22 & 0.73 $\pm$ 0.01 & 166 $\pm$ 1 & 5.95 & 2.540 $\pm$ 0.009 & 115.7 $\pm$ 0.2 & 6.36 & 1.356 $\pm$ 0.008 & -125.1 $\pm$ 0.3 \\
    \hline
    \end{tabular*}
    \begin{tablenotes}
	\item[a] Fitting errors of $v_0$ are less than 0.1 \%.
	\end{tablenotes}
\end{threeparttable}
\end{table*}
%\end{longrotatetable}
% ---

% Peak velocity measurement
\subsection{Peak Velocity Measurements \label{subsec:ana_vpeak}}
The small-scale maps of L1527 IRS and TMC-1A exhibit velocity gradients, which likely trace rotational motion of the disks and envelopes, although such a velocity gradient due to rotational motion of the disk and envelope is not very clear in the IRAS 15398$-$3359 because of the small rotational velocity compared to those in other sources \citep{Yen:2017aa}. On the other hand, velocity gradients are less clear and more perturbed in the large-scale maps of all three sources. In this subsection, we measure the peak velocity as a function of radius along the disk major axis and examine its radial dependence to investigate how the nature of the velocity structure changes with radius.

For this purpose, position-velocity (PV) diagrams cut along the disk major axis were produced from both small- and large-scale maps. The velocity component along a disk major axis mainly reflects rotational velocity and is less affected by outflows. Therefore, PV cut along the disk major axis is reasonable to trace rotation motion at inner radii and investigate how it changes with changing radii. Position angles of 140$^\circ$, 0$^\circ$ and 70$^\circ$ are adopted for the disk major axes of IRAS 15398$-$3359, L1527 IRS and TMC-1A, respectively, based on previous observations of their disks \citep{Tobin:2012aa, Oya:2014aa, Ohashi:2014aa, Aso:2015aa, Aso:2017ab, Yen:2017aa}. The width of the PV cuts is one pixel width ($\sim$1$/10$ beam size), which means no averaging pixels perpendicular to the slices.

\begin{figure*}[htbp]
\centering
\includegraphics[width=2\columnwidth]{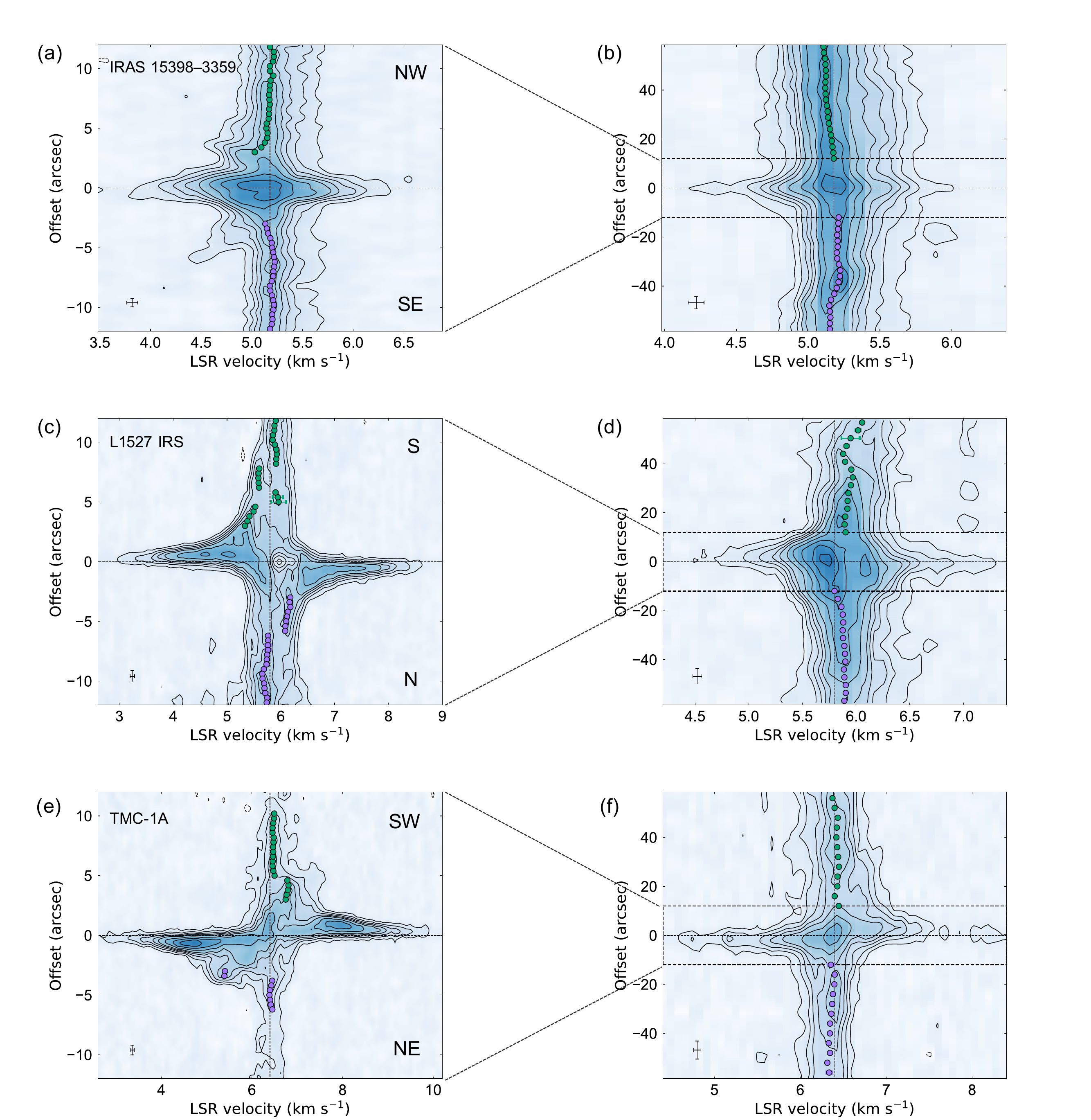}
\caption{PV diagrams of the C$^{18}$O 2--1 emissions cut along the disk major axes. Contour levels are from 3$\sigma$ to 12$\sigma$ in steps of 3$\sigma$ and from 12$\sigma$ to 60$\sigma$ in steps of 6$\sigma$, where 1$\sigma$ is rms noise summarized in Table \ref{tab:mapsummary}. Green and purple markers show the peak velocities measured on opposite sides of the protostars through the Gaussian fitting. Vertical dashed lines denote the systemic velocities. Vertical and horizontal bars in the bottom-left corners indicate the spatial and velocity resolutions, respectively. Scales of the velocity axis are different in the right and left columns.}
\label{fig:pvdiagrams}
\end{figure*}

The generated PV diagrams are presented in Figure \ref{fig:pvdiagrams}. Small-scale PV diagrams of all the three protostars, shown in Figure \ref{fig:pvdiagrams}(a), (c) and (e), exhibit high-velocity components at small offsets within $\sim\pm\ang[angle-symbol-over-decimal]{;;2}$. The emission peak at each position appears at a higher velocity with respect to the systemic velocity as the offset decreases at offsets of $\sim\ang[angle-symbol-over-decimal]{;;3}$--$\ang[angle-symbol-over-decimal]{;;8}$ in L1527 IRS and TMC-1A, although this feature is not very clear in IRAS 15398$-$3359. These velocity structures can be interpreted as the differential rotation of the envelope, where the rotational velocity increases with decreasing radius. In addition to these velocity structures, the small-scale PV diagram of IRAS 15398$-$3359 shows an emission whose velocity appears to linearly increase with the radius at offsets of $\sim\ang[angle-symbol-over-decimal]{;;-6}$ to $\ang[angle-symbol-over-decimal]{;;-3}$. This velocity structure was also reported in previous ALMA observations by \cite{Okoda:2021aa}, who suggests the velocity component originates from the secondary outflow launched in a direction perpendicular to the primary outflow. On the other hand, velocity gradient is less clear and feature of differential rotation is not seen at offsets of $\gtrsim\ang[angle-symbol-over-decimal]{;;12}$ and $\lesssim\ang[angle-symbol-over-decimal]{;;12}$ in the large-scale PV diagrams presented in Figure \ref{fig:pvdiagrams}(b), (d) and (f). Moreover, the emission peak at each position at those offsets appears at redshifted velocity on both the north and south sides in L1527 IRS, which would not be explained by rotational motion.
% A similar velocity structure, whose the velocity appears to linearly increase with radius, is seen in TMC-1A in Figure \ref{fig:pvdiagrams}(e) at offset of $\sim\ang[angle-symbol-over-decimal]{;;-5}$ to $\ang[angle-symbol-over-decimal]{;;5}$. just before "on the other hand".

% ---------- Figure: analysis ---------
% PV diagrams
% ----------------------------------

In order to measure the peak velocity as a function of radius, the method used in \cite{Sai:2020aa, Sai:2022aa} is adopted, i.e., a Gaussian function is fitted to the spectrum at each offset in the PV diagrams. The offset coordinates are sampled in steps of a half of the beam size, and the spectrum measured on a single pixel at a given offset is used for the fitting. The fitting is performed only to the spectrum whose detected peak intensity is above 6$\sigma$. The $\pm2$ velocity channels around the intensity maximum of the spectrum are used for the fitting to better trace the peak velocity of the non-Gaussian shape spectrum \citep[][]{Sai:2020aa}. The peak velocities are measured at offset ranges of $\ang[angle-symbol-over-decimal]{;;3}$ to $\ang[angle-symbol-over-decimal]{;;12}$ and $-\ang[angle-symbol-over-decimal]{;;12}$ to $-\ang[angle-symbol-over-decimal]{;;3}$ using the small-scale PV diagrams. The offset range of $-3''$ to $3''$ are excluded from the fitting, since the emission peaks appear at low velocities and the peak velocities no longer trace the feature of differential rotation within the offset range. The peak velocities outside $\ang[angle-symbol-over-decimal]{;;12}$ are measured using the large-scale PV diagrams. The error of the measured peak velocity is derived as the fitting error.

The measured data points are overlaid on the PV diagram in Figure \ref{fig:pvdiagrams}. Figure \ref{fig:pvdiagrams}(a) shows that the peak velocities of IRAS 15398$-$3359 measured on the small-scale are overall blueshifted on the northwest side and redshifted on the southeast side of the protostar with respect to $\vsys$, which may suggest rotational motion despite the very small measured velocity. Peak velocities measured with the large-scale PV diagram in Figure \ref{fig:pvdiagrams}(b) show a similar feature, while the peak velocity at offsets larger than $40''$ on the southeast side is blueshifted. The peak velocity is further blueshifted as radius increases on the northwest side. Figure \ref{fig:pvdiagrams}(c) shows that the peak velocities measured at the small-scale for L1527 IRS roughly trace the feature of differential rotation within offset ranges of $\ang[angle-symbol-over-decimal]{;;3}$ to $\ang[angle-symbol-over-decimal]{;;5}$ on the south side and $-\ang[angle-symbol-over-decimal]{;;7}$ to $-\ang[angle-symbol-over-decimal]{;;3}$ on the north side, while the peak velocities show abrupt changes across the systemic velocity outside of these offsets. The peak velocities measured with the large-scale PV diagram shown in Figure \ref{fig:pvdiagrams}(d), on the other hand, are redshifted on both the north and south sides. Moreover, they show more redshifted velocities with respect to the systemic velocity at larger radii. The peak velocities measured for TMC-1A also show abrupt changes at a offset of $\sim-\ang[angle-symbol-over-decimal]{;;4}$ on the northeast side and $\sim\ang[angle-symbol-over-decimal]{;;5}$ on the southwest side across the systemic velocity in the small-scale PV diagram, as presented in Figure \ref{fig:pvdiagrams}(e). In the large-scale PV diagram shown in Figure \ref{fig:pvdiagrams}(f), most of the peak velocities on the northeast and southwest sides are blueshifted and redshifted, respectively, with respect to the systemic velocity.

\begin{figure*}[htbp]
\centering
\includegraphics[width=2\columnwidth]{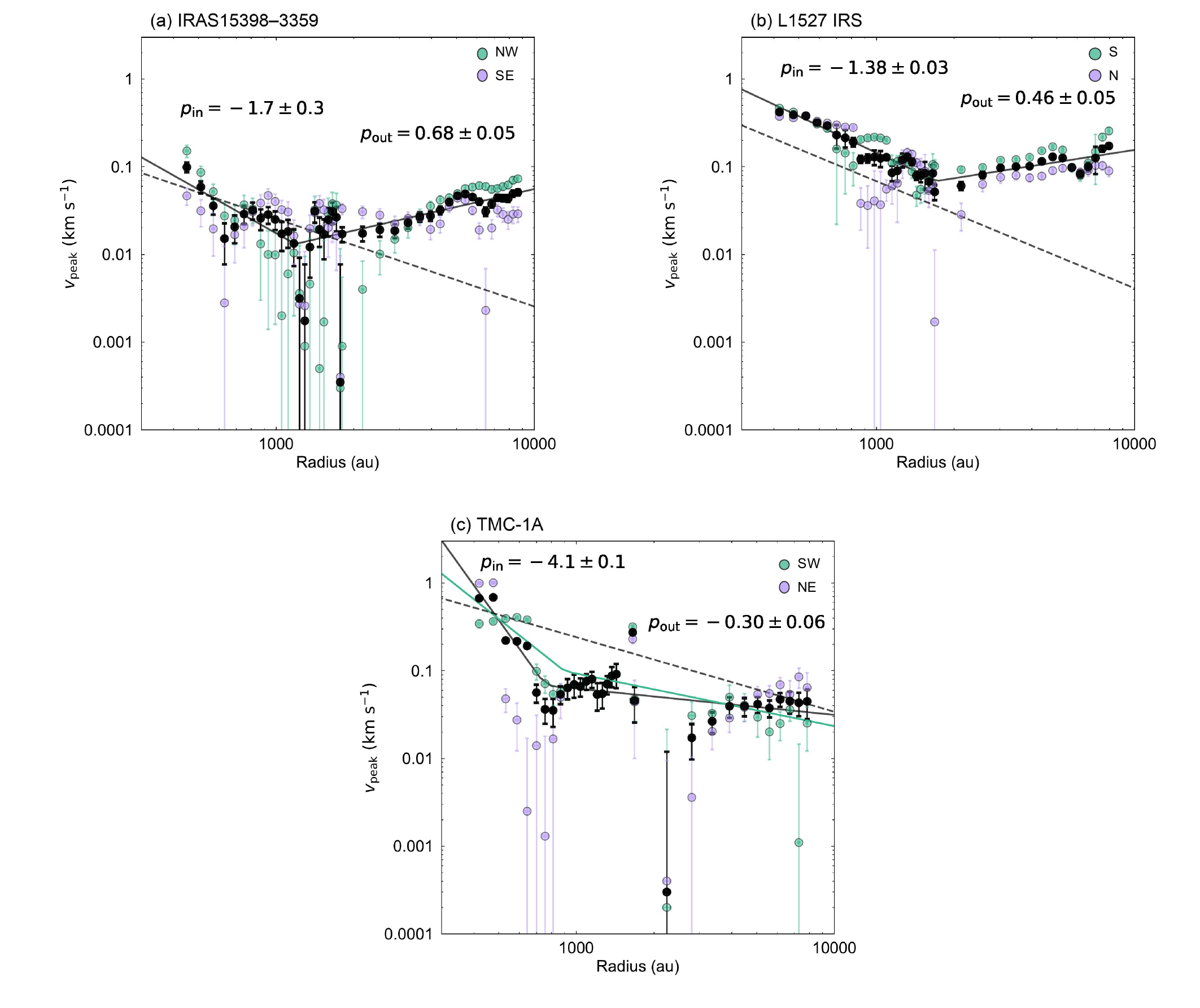}
\caption{Radial profiles of the peak velocities measured with the PV diagrams. Green and purple circles show the data points measured on opposite sides of the protostars and the black circles represent the averaged value at each radius. The dashed lines are the extrapolated best-fit power-law functions measured at $\sim$100 au in previous works \citep{Aso:2015aa, Aso:2017ab, Yen:2017aa}. The black, solid lines show the double power-law functions with the best-fit parameters for the averaged peak velocities in this work. The green, solid line shows the double-power law function with the best-fit parameter for the peak velocity measured on the southwest side of TMC-1A.}
\label{fig:vprof_all}
\end{figure*}

Figure \ref{fig:vprof_all} presents the measured peak velocities in $\log r$-$\log v$ diagrams in the form of relative velocity, $\vpeak=|\vlsr - \vsys|$, to investigate their radial dependence. In addition to the measured data points represented in colored circles, averaged peak velocities, which are calculated by averaging the velocities obtained at a given offset assuming an azimuthally symmetric velocity structure, are also plotted in black circles in the diagrams.

The radial profile of the averaged peak velocity for IRAS 15398$-$3359 shown in Figure \ref{fig:vprof_all}(a), overall, exhibits a systematic trend as a function of radius: the averaged peak velocities decrease and then increase with increasing radius with a transition around a radius of $\sim$1200 au. In addition, the radial profile shows small bumps at radii of $\sim$900 au and $\sim$1500 au. The averaged peak velocities of L1527 IRS presented in Figure \ref{fig:vprof_all}(b) also decrease and then increase with increasing radius with a transition radius of $\sim$1700 au. Similar radial velocity profiles with such two regimes have been observed for envelopes around 12 Class 0 protostars by calculating an averaged profile among the sources \citep{Gaudel:2020aa} and also in the envelope around the Class I protostar L1489 IRS \citep[][]{Sai:2022aa}. TMC-1A also exhibits a break at a radius of $\sim$800 au in the radial profile of the averaged peak velocity, as shown in Figure \ref{fig:vprof_all}(c). However, the measured peak velocities are more scattered than those of the other two sources. Moreover, the peak velocities measured on the southwest and northeast sides of the protostar are mostly inconsistent with each other at small radii of $\lesssim$800 au.

To characterize the radial profiles showing two regimes, the following double power-law function was fitted to the averaged peak velocities with the MCMC code \texttt{emcee} \citep[][]{Foreman-Mackey:2013aa}:
\begin{align}
    \vpeak =
    \begin{cases}
	\vbreak \left( \frac{r}{r_\mathrm{break}} \right)^{p_\mathrm{in}}
	 & \left(r \leq r_\mathrm{break} \right) \\
	\vbreak \left( \frac{r}{r_\mathrm{break}} \right)^{p_\mathrm{out}}
	 & \left(r > r_\mathrm{break} \right)
	\end{cases},
\end{align}
where $\rbreak$ is the break radius where the power-law the function changes. In the double power-law fitting, ($\vbreak$, $r_\mathrm{break}$, $p_\mathrm{in}$, $p_\mathrm{out}$) were set as free parameters and their parameter spaces were searched. The best-fit parameters and their errors are obtained as the mean and standard deviation of the posterior probability distribution, respectively. The results of the fitting are summarized in Table \ref{tab:plawfit}.

% TMC-1A
For TMC-1A, it could not be appropriate to assume an azimuthally symmetric envelope and use the averaged peak velocity, as previous observations have suggested an asymmetric molecular gas distribution on $\sim$100--500 au scales\citep{Sakai:2016aa, Harsono:2021aa}. Hence, we performed fittings using the double-power law function to the velocities measured on each side of the protostar separately without averaging as well as to the averaged peak velocities. While the fitting for the data points on the southwest side is converged, we could not obtain any solution for the data points on the northeast side with the double-power law function. These fitting results are also summarized in Table \ref{tab:plawfit}.

% --------- Table ----------
% vpeak fit
\begin{table*}[htbp]
\begin{threeparttable}
    \centering
    \caption{Best-fit parameters of the double-power law fitting to the measured peak velocities}
    \label{tab:plawfit}
    \begin{tabular*}{2\columnwidth}{@{\extracolsep{\fill}}lcccc}
    \hline
    \hline
     Source & $\vbreak$ ($\kmps$) & $\rbreak$ (au) & $p_\mathrm{in}$ & $p_\mathrm{out}$ \\
    \hline
    IRAS 15398$-$3359 & $0.013\pm 0.001$ & $1180 \pm 90$ & $-1.7 \pm 0.3$ & $0.68 \pm 0.05$ \\
    L1527 IRS & $0.069 \pm 0.003$ & $1700 \pm 50$ & $-1.38 \pm 0.03$ & $0.46 \pm 0.05$ \\
    TMC-1A & $0.068 \pm 0.005$ & $760 \pm 10$ & $-4.1 \pm 0.1$ & $-0.30 \pm 0.06$ \\
    TMC-1A (SW side only) & $0.099 \pm 0.007$ & $900 \pm 30$ & $-2.33 \pm 0.09$ & $-0.60 \pm 0.07$ \\
    \hline
    \end{tabular*}
    \begin{tablenotes}
    \item {\footnotesize The first three rows show results of the fitting to the averaged peak velocities, and the last row shows a result of the fitting to the peak velocities measured on the southwest side of TMC-1A.}
\end{tablenotes}
\end{threeparttable}
\end{table*}
% ---------------------

% ------------ Figure: analysis -------------
% delta v profiles
% ---------------------------------------------

\section{Discussion} \label{sec:discussion}
\subsection{Nature of the Radial Profiles of the Peak Velocity \label{subsec:discussion_vpeakprof}}
\subsubsection{IRAS 15398$-$3359}
% brief summary of our result and previous work
The power-law index within the break radius is $-1.7 \pm 0.3$ for IRAS 15398$-$3359. A similar measurement was made by \cite{Yen:2017aa}, who estimated a power-law index to be $\sim-$1.0 within a radius of $\sim$100 au in C$^{18}$O $J=2$--1 and interpreted the power-law index as rotational motion of the infalling envelope conserving specific angular momentum.

% possible explanation of different slopes
The measured power-law index in the current work is smaller than that derived in the previous work, although the fitting uncertainty is also large. This difference could be due to contamination of gas motion on large spatial scales, as all velocity components on different spatial scales along the line of sight are included in the current maps. The peak velocity of the spectrum may be more strongly affected by the large-scale gas motion around the centroid velocity of the foreground gas. This stronger contamination of the large-scale gas motion at particular velocity would change the slope of the radial profile of the peak velocity. Thus, the measured power-law index of $-1.7\pm0.3$ could be interpreted as indicating a rotational motion of the envelope with a degree of contamination of the large-scale gas motion.
% further components not likely coming from pure rotation.

It is suggested that the large-scale gas is turbulent in IRAS 15398$-$3359, as discussed in Section \ref{subsec:vfield_tb}. The contamination of the turbulent, large-scale gas motion could be also a cause of the two small bumps seen in the radial profile of the peak velocity at radius of $\sim$900 au and $\sim$1500 au in Figure \ref{fig:vprof_all}(a). The bumps have a width of $\sim$400 au, and the deviation of the peak velocities around the bumps is $\sim$0.01 $\kmps$. Our analysis suggests that the turbulent motion in IRAS 15398–3359 is characterized by a relation of $\delta v = 10^{-1.69} (\tau/1000~\au)^{0.6}$, where $\delta v$ is the velocity deviation and $\tau$ is the spatial scale (see Section \ref{subsec:vfield_tb} for more detail). This relation predicts velocity diﬀerence of $\sim$0.01 $\kmps$ on a scale of 400 au, which matches the deviation of the peak velocities around the bumps.

One might wonder whether the secondary outflow component suggested in a previous work \citep[][]{Okoda:2021aa} affects our analysis, as it is present on the southeast side in the cut direction of the PV diagram. The secondary outflow component appears at velocities of 4.5--5.0 $\kmps$ within an offset range of $-6''$ to $-3''$, while the peak velocities measured within the offset range are around 5.18 $\kmps$. Thus, our measurements mainly trace different velocity components from the secondary outflow. On the other hand, the secondary outflow could skew the overall shape of the spectrum and cause a shift of the peak velocity. A few data points around $\pm$3$''$ offsets ($\sim$450 au) in the PV diagram in Figure \ref{fig:pvdiagrams}(a) have blueshifted velocity on both sides of the protostar, while most of data points have blueshifted and redshifted velocities on the northwest and southeast sides, respectively, as expected for its envelope rotation \citep{Oya:2014aa, Yen:2017aa}. Those velocity components that are not expected for a pure rotational motion on the southeast side could be influenced by the secondary outflow.

The power-law index of the radial profile outside the break radius, on the other hand, is 0.68$\pm$0.05 and significantly different from that inside the break radius. This increase of the peak velocity with
radius is consistent with previous measurements of velocity structures within dense cores on scales of $\sim$1000--10,000 au, i.e., $v \sim j/r \propto r^{0.6\operatorname{--}0.8}$ \citep{Goodman:1993aa, Caselli:2002aa, Pineda:2019aa, Gaudel:2020aa}. The origin of the velocity structures outside the break radius is further discussed in Section \ref{subsec:vfield_tb}.

% L1527 IRS
\subsubsection{L1527 IRS}
The power-law index within the break radius is $-1.38 \pm 0.03$ for L1527 IRS. \cite{Aso:2017ab} have also measured a radial profile of the gas velocity for L1527 IRS within a radius of $\sim$100 au, giving a power law index of $-1.22\pm0.04$, which roughly agrees with our measurement. Although the indices of the two measurements are consistent with each other, the measured velocities are not consistent between two measurements, as shown in Figure \ref{fig:vprof_all}(b). The peak velocities measured in the current work are a few time larger than those expected from the extrapolation of the radial profile measured by \cite{Aso:2017ab} while assuming the same systemic velocity. Such a discrepancy could be caused by contamination of infalling motions due to a relatively lower angular resolution of our measurements, as discussed by \cite{Sai:2022aa}. \cite{Gaudel:2020aa} have also measured the line-of-sight velocity as a function of radius at radii of 50--8000 au in L1527 IRS, and fitted a double-power law function to its apparent specific angular momentum $|j_\mathrm{app}|=r \times v$. They report a break radius of 1320$\pm$260, and power-law indices of the radial dependence of the apparent specific angular momentum of $-0.14\pm0.03$ and $1.3\pm0.2$ inside and outside the break radius (see their Table F.2), corresponding to $p\sim-1.14$ and $0.3$ of $v \propto r^{p}$, respectively. The break radius and power-law indices they reported are roughly consistent with the values derived in the current work despite the difference in data sets and fitting methods.

The power-law index we measured within the break radius is smaller than $-1$, which is expected for rotational motion of the envelope conserving specific angular momentum. We also see abrupt changes of the peak velocity seen in the PV diagram (Figure \ref{fig:pvdiagrams}), as discussed in Section \ref{subsec:ana_vpeak}. Importantly, the radii where the abrupt changes of the peak velocities occur are different between the north and south sides and are smaller than the break radius, suggesting that the entire motions within the break radius cannot be explained by simple rotational motions. A possible explanation of these velocity structures is contamination of gas motion on larger scales, as discussed for IRAS 15398$-$3359 in the previous subsection. The actual spatial scale where the velocity structure follows the rotational motion could be, thus, smaller than the estimated break radius, as \cite{Gaudel:2020aa} reported a smaller break radius of 1320 au.

Similarly to IRAS 15398$-$3359, the power-law index outside the break radius is very different from that inside the break radius for L1527 IRS. The interpretation of the velocity structure outside the break radius is discussed in more detail in Section \ref{subsec:vfield_tb}.

\subsubsection{TMC-1A}
The radial profile of the peak velocity for TMC-1A is very different from those of the two other protostars described above. The power inside of the break radius, $-4.1 \pm -0.1$ for the averaged peak velocity and $-2.33 \pm -0.09$ on the southwest side of the protostar, is much smaller than in IRAS 15398-3559 and L1527 IRS. These values also varies greatly from previous measurements of the rotational velocity made by \cite{Aso:2015aa}, reporting the power-law index of $-0.86$ within a radius of $\sim$100 au. Moreover, the peak velocities measured on the southwest and northeast sides of the protostar are mostly inconsistent with each other within the break radius. These facts suggest that the measured peak velocity in TMC-1A in the current work does not seem to trace the rotational velocity of the infalling envelope even within the break radius. One possible reason for this is an asymmetric structure of its infalling envelope. An asymmetric molecular distribution has been reported on $\sim$100--500 au scales in TMC-1A in a previous work \citep{Sakai:2016aa, Harsono:2021aa}, and it is explained with an infalling streamer-like structure caused by an infalling cloudlet \citep[][]{Hanawa:2022aa}. In such a case where infalling gas is asymmetric, the line-of-sight velocity along the rotational major axis can represent infalling velocity rather than rotational velocity. The projected infalling velocity at a radius of 400 au on the sky plane is a maximum of $\sim$1.9 $\kmps$ according to the protostellar mass of 0.68 $\Msun$ \citep[][]{Aso:2015aa}. Thus, it would be possible that inflows produce the peak velocities of $\sim$0.4--1 $\kmps$ around a radius of 400 au depending on the projection angle.

The power outside of the break radius, $-0.30 \pm 0.06$ or $-0.60\pm 0.07$, is also significantly different from those for the other two sources. There seems to be a less clear velocity gradient compared to the other two sources. One possibility is that foreground gas of TMC-1A has little motions and its velocity component close to the systemic velocity is dominant in the current maps due to integration along the line-of-sight. We also note that the peak velocities outside the break radius are small, particularly at radii larger than 2000 au on the southwest side; most of them are less than 0.03 $\kmps$ and have relatively large uncertainties of $\sim$50\% of the measured peak velocities. Thus, the peak velocities measured at the large radii could be close to the accuracy limit.

\subsection{Origin of the Velocity Structure Outside the Break Radius
\label{subsec:vfield_tb}}
The velocity structures outside the break radii in IRAS 15398$-$3359 and L1527 IRS are not likely due to simple rotational motion. A possible origin of the incoherent velocity structures on the large scale is the non-axisymmetric collapse of the dense core. A numerical simulation performed by \cite{Verliat:2020aa} demonstrates that such non-axisymmetric collapse generates local rotational motion, and thus velocity gradients appear at scales of hundreds to thousands of au. In their simulation where the initial dense core has density perturbations but no motion, the specific angular momentum calculated using projected velocity as $j=r^2 \Omega$, where $\Omega$ is the two dimensional velocity gradient measured with simulated observational maps, is a few times $10^{-4}~\kmps$ pc at a radius of 1000 au. This value agrees with values of $\sim$10$^{-4}$--$10^{-3}~\kmps$ pc calculated from the current observations as $j=r \times \vpeak$. On the other hand, the specific angular momentum calculated in their simulation at larger radius from 2000 to 8000 au is (2--6) $\times$10$^{-4}~\kmps$ pc, which are smaller by an order of magnitude at a maximum than (5--80)$\times$10$^{-4}~\kmps$ pc calculated from our observations within the same radius range. This difference may arise because their calculation simulates an isolated self-standing core with no motion and that is not embedded in large scale structures. The angular momentum at larger scales could also depend on the degree of the initial density perturbation in the simulations. As their study focused on the inner envelope scales, further models treating both the large scale structures and density perturbations are needed to investigate the velocity structures caused by non-axisymmetric collapse on scales of $\geq$2000 au in more detail.

Another possibility to explain the kinematic structure outside the break radius is turbulence motion. It is suggested that scale-dependent turbulence in dense cores can produce a velocity gradient on subparsec scales in projected maps \citep[][]{Burkert:2000aa}. The turbulence in molecular clouds is known to follow the Larson's law, $\Delta v \propto L^{\sim0.5}$, where $L$ is the cloud size and $\Delta v$ is the velocity deviation over the cloud \citep{Larson:1969aa, Heyer:2004aa, McKee:2007aa}. The relation of $\vpeak \propto r^{0.5\operatorname{--}0.6}$ measured in IRAS 15398$-$3359 and L1527 IRS is similar to the scaling law, as some previous works pointed out the similarity of the Larson's law and the relation of $J/M\propto R^{1.6}$ (corresponding to $v \propto r^{0.6}$) measured in dense cores \citep[][]{Goodman:1993aa, Ohashi:1997ab, Tatematsu:2016aa, Gaudel:2020aa}.

To examine whether the velocity structures on a large scale originate from turbulent motion, spatial correlations of the velocity deviation are examined with the larges-scale moment 1 maps (i.e., the centroid velocity) through the second-order structure function (SF) and the autocorrelation function (ACF). The slope of the spatial correlation of the velocity deviation, $\gamma_\mathrm{2D}$, where $\delta v \propto \tau^{\gamma_\mathrm{2D}}$, is compared with that observed in molecular clouds and predictions of turbulence models. Note that the slope measured from the centroid velocity is denoted as $\gamma_\mathrm{2D}$ since it does not necessarily equal to the slope $\gamma_\mathrm{3D}$, which is measured in three dimensions, as discussed later. The large-scale maps include velocity components of infalling envelopes and outflows, as was reviewed in Section \ref{sec:results}. To exclude these components, one-dimensional slices of the large-scale moment 1 maps along the disk major axis are used with masks excluding the velocity components inside the break radii. The one-dimensional slices of the moment 1 maps are sampled with a half of the beam size along the cut direction. The effect of masking on the slope $\gamma_{2D}$ is examined using model turbulent-velocity fields in Appendix \ref{sec:app_tbmodel}, demonstrating that masking does not affect the value of the slope for turbulent-velocity field significantly.

The discrete, one-dimensional SF and ACF can be defined, respectively, as follows \citep{Miesch:1994aa};
\begin{align}
    S(\tau) = N(\tau)^{-1} \sum_x [v_\mathrm{c}(x) -v_\mathrm{c}(x+\tau)]^2,
\end{align}
and
\begin{align}
    C(\tau) = \sigma_{v_\mathrm{c}}^{-2}N(\tau)^{-1} \sum_x [v_\mathrm{c}(x) - \mu][v_\mathrm{c}(x+\tau) - \mu],
\end{align}
where $\mu$ is the mean centroid velocity and defined as
\begin{align}
    \mu \equiv \frac{\sum v_\mathrm{c}(x)}{N_\mathrm{tot}},
\end{align}
and $ \sigma_{v_\mathrm{c}}^2$ is the variance of the centroid velocity and defined as
\begin{align}
    \sigma_{v_\mathrm{c}}^2 \equiv \frac{\sum [v_\mathrm{c}(x) - \mu]^2}{N_\mathrm{tot}}.
\end{align}
Here, $S(\tau)$ is the SF, $C(\tau)$ is the ACF, $x$ is the coordinate in the one-dimentional slice of the moment 1 map, $v_\mathrm{c}$ is the centroid velocity, $\tau$ is the spatial separation between two pixels and $N$ is the number of pixels that used to calculate correlations. The spatial correlation of the velocity deviation is obtained as $\delta v(\tau)=\sqrt{S(\tau)}$. ACFs provide the information of the effective, largest scale of turbulence $\tau_0$, where $C(\tau_0) = 0$ \citep{Miesch:1994aa, Miville-Deschenes:1995aa}. Because the velocity deviation is decorrelated with the spatial scale at scales larger than $\tau_0$, the spatial correlation of the velocity deviation is examined within $\tau_0$, following previous studies \citep[e.g.,][]{Miesch:1994aa}. Uncertainties of ACFs and SFs are estimated through the error propagation.

% ------------ Figure: analysis -------------
% ACFs
\begin{figure*}[tbhp]
\centering
\includegraphics[width=2\columnwidth]{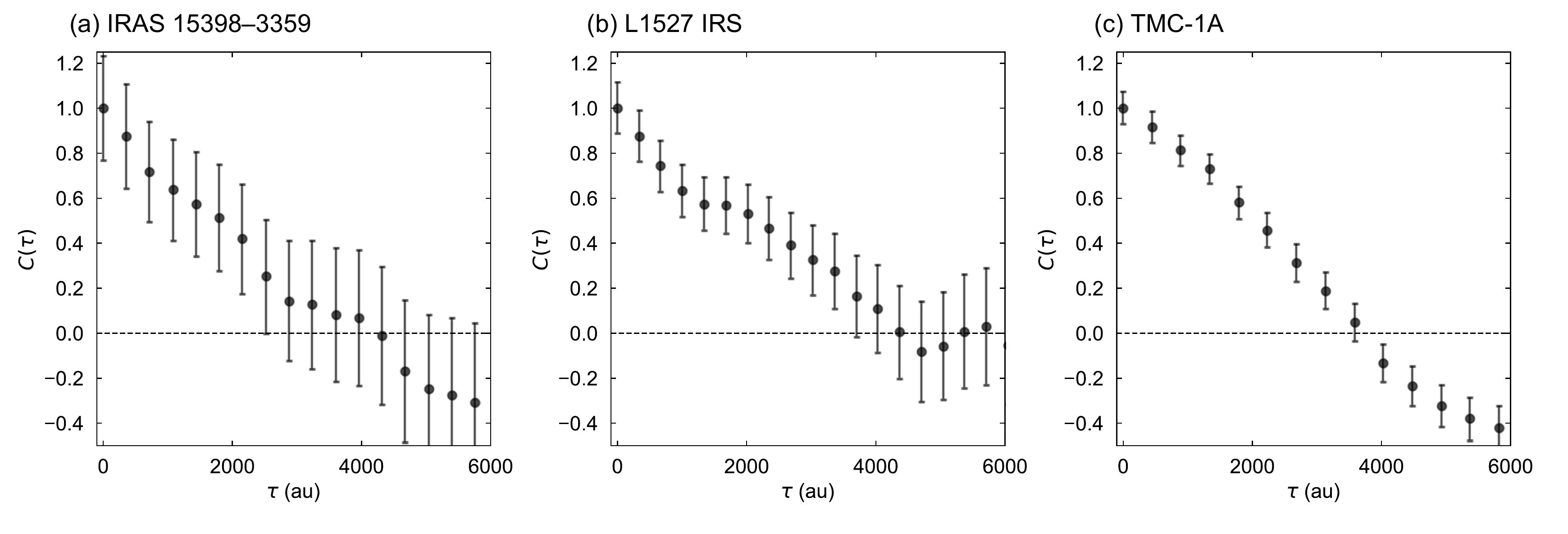}
\caption{ACFs calculated from the one-dimensional slices of the large-scale moment 1 maps.}
\label{fig:acf_all}
\end{figure*}

% SFs
\begin{figure*}[htbp]
\centering
\includegraphics[width=2\columnwidth]{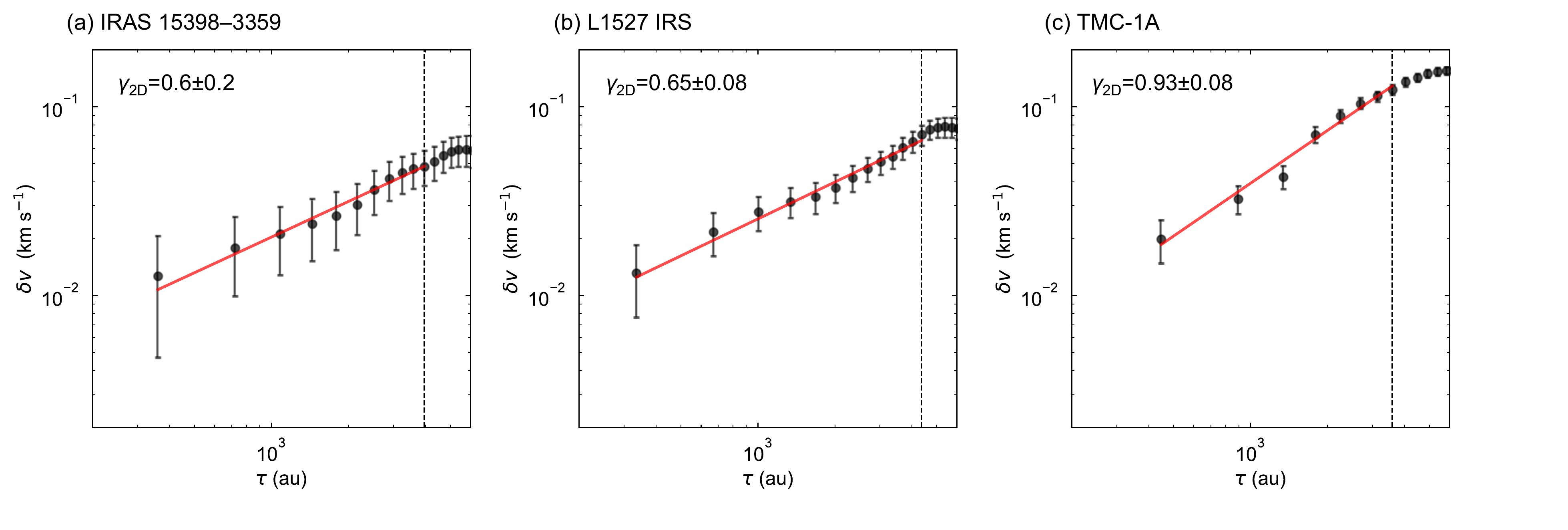}
\caption{Spatial correlations of the velocity deviation around the three protostars. Red solid lines indicate the best-fit power-law relations. Vertical dashed lines show $\tau_0$, within which the fitting was performed. }
\label{fig:delv_all}
\end{figure*}
% ---------------------------------------------

Figure \ref{fig:acf_all} presents ACFs calculated from the one-dimensional slices. The ACFs of IRAS 15398$-$3359 and L1527 IRS almost monotonically decrease until they reach zero, providing $\tau_0$ of $\sim$4000 au and 4400 au, respectively. Calculated $\delta v (\tau)$ for the two sources are presented in Figure \ref{fig:delv_all}(a) and (b). The spatial scale and velocity deviation exhibit a strong positive correlation within $\tau_0$. A linear function is fitted to $\delta v (\tau)$ in the log-log space through $\chi^2$ fitting. The best-fit power-law relations is
\begin{align}
    \delta v = 10^{-1.69 \pm 0.08} \left( \frac{\tau}{1000~\au} \right)^{0.6\pm0.2} ~ (\kmps)
\end{align}
for IRAS 15398$-$3359, and
\begin{align}
    \delta v = 10^{-1.59 \pm 0.04} \left( \frac{\tau}{1000~\au} \right)^{0.65\pm0.08} ~ (\kmps)
\end{align}
for L1527 IRS. We also performed the same analysis using one-dimensional cuts with position angles that deviate from the position angle of the major axis within $\pm10^\circ$ and found that the uncertainty of the slope associated with the cut direction is $\sim0.1$ within the range of the position angle.

We also calculated a spatial correlation of the velocity deviation for TMC-1A, although the break radius of TMC-1A seems not to represent the transition between the infalling envelope with a constant specific angular momentum and outside. The ACF for TMC-1A also monotonically decreases with the spatial scale, providing $\tau_0$ of 3600 au, as shown in Figure \ref{fig:acf_all}(c). The calculated $\delta v(\tau)$ for TMC-1A has a steeper profile compared to the other two sources, as seen in Figure \ref{fig:delv_all}(c). As such, the best-fit power-law function is $\delta v = 10^{-1.41\pm0.03} (\tau/1000~\au)^{0.93\pm0.08}$. All fitting results are summarized in Table \ref{tab:param_delvfit}.

% Comparison to measurements in molecular clouds
Spatial correlations of the velocity deviation (or SFs) have been studied in molecular clouds using centroid velocity maps. An earlier work by \cite{Miesch:1994aa} reported a wide range of $\gamma_\mathrm{2D}$ from 0.23 to 0.79, with a mean and standard deviation of 0.50 and 0.16, respectively, at spatial scales of $\sim$0.1--1 pc in 14 molecular clouds. \cite{Ossenkopf:2002aa} applied the $\Delta$-variance method, which is similar to the SF but calculates the average variance on a certain spatial scale with the filtered map, to the centroid maps of the Polaris Flare cloud, suggesting that $\gamma_\mathrm{2D}\sim$0.32--0.68 within a scale of $\sim$1500 au to 0.75 pc. The slope $\gamma_\mathrm{2D}$ derived for IRAS 15398$-$3359 and L1527 IRS in the current work is similar to the larger values derived in molecular clouds mentioned above, although that of TMC-1A is larger than the values reported in the previous works.

% --------- Table ----------
% vpeak fit
\begin{table}[thbp]
\begin{threeparttable}
    \centering
    \caption{Results of the fitting to the spatial correlation of the velocity deviation.}
    \label{tab:param_delvfit}
    \begin{tabular*}{\columnwidth}{@{\extracolsep{\fill}}lccc}
    \hline
    \hline
     Source & $r_\mathrm{mask}$ (au) & $q$\tnote{1}  & $\gamma_\mathrm{2D}$\tnote{1}  \\
    \hline
    IRAS 15398$-$3359 & 1200 & $-1.69\pm0.08$ & $0.6\pm0.2$ \\
     & 1700 & $-1.69\pm0.08$ & $0.6\pm0.2$ \\
     & 2400 & $-1.68\pm0.08$ & $0.5\pm0.3$ \\
    L1527 IRS & 1700 & $-1.59\pm0.04$ & $0.65\pm0.08$ \\
     & 2400 & $-1.58\pm0.04$ & $0.6\pm0.1$ \\
    TMC-1A & 800 & $-1.41\pm0.03$ & $0.93\pm0.08$ \\
     & 1700 & $-1.52\pm0.04$ & $0.83\pm0.09$ \\
     & 2400 & $-1.50\pm0.04$ & $0.74\pm0.09$ \\
    \hline
    \end{tabular*}
    \begin{tablenotes}
	\item[1] $\delta v(\tau)=10^{q} (\tau/\mathrm{1000~\au})^{\gamma_\mathrm{2D}}$
	\end{tablenotes}
\end{threeparttable}

\end{table}
% ---------------------

% Comparison to theoretical works
In order to compare the slopes derived using the centroid velocity with theoretical predictions, the projection effect has to be considered. $\gamma_\mathrm{2D}$ measured with centroid velocity maps is expected to be larger than $\gamma_\mathrm{3D}$ measured in the three dimension, representing true nature of turbulence, since the velocity fluctuation along the line-of-sight averages out in centroid velocity maps \citep{Odell:1987aa, Miville-Deschenes:2003aa, Brunt:2003ab}. \cite{Brunt:2004aa} studied this effect using HD and MHD simulations in detail, taking into account density inhomogeneity, and derived an empirical equation for the conversion from $\gamma_\mathrm{2D}$ to $\gamma_\mathrm{3D}$:
\begin{align}
    \gamma_\mathrm{2D} = \gamma_\mathrm{3D} + 0.5 + \delta \kappa/2,
    \label{eq:gamma2dto3d}
\end{align}
where $\delta \kappa$ is an empirically derived constant. They suggested that, when turbulent energy is being injected (driven turbulence), $\delta \kappa$ ranges from $-1.5$ to $-0.5$ and  statistically approaches $\sim -1$ (i.e., $\gamma_\mathrm{2D}\sim \gamma_\mathrm{3D}$). This is because density fluctuation causes additional small-scale fluctuation in the centroid velocity through intensity weighting and the projection-smoothing effect is canceled out. The coefficient $\delta \kappa$, on the other hand, decreases to zero when injection of turbulent energy is turned off and turbulence is decaying (decaying turbulence). We, therefore, estimate the possible $\gamma_\mathrm{2D}$ values for the both cases of driven turbulence ($\delta \kappa\sim -1.5$--$-0.5$) and decaying turbulence ($\delta \kappa=0$) with Equation (\ref{eq:gamma2dto3d}).

Widely accepted theoretical models of isotropic turbulence, Kolmogorov turbulence and Burgers turbulence, expect $\gamma_\mathrm{3D}=1/3$ for incompressible fluids \citep{Kolmogorov:1941aa}, and $\gamma_\mathrm{3D} = 1/2$ for compressible supersonic turbulence \citep{Burgers:1974aa}. In these models, $\gamma_\mathrm{2D}$ is expected to be $\sim$0.83–1.0 for the case of decaying turbulence, which is much larger than those measured for IRAS 15398$-$3359 and L1527 IRS. On the other hand, the possible ranges of $\gamma_\mathrm{2D}$ for the case of driven turbulence are $\sim$0.08--0.75. The slopes derived in IRAS 15398$-$3359 and L1527 IRS are consistent with this range. A simulation work suggests that outflows associated with low-mass protostars can drive and maintain turbulence in an isolated core over an entire lifetime of Class 0 protostars of $\sim$10$^5$ yrs \citep{Offner:2014ab,Offner:2017aa}. Hence, it could be possible that turbulence is being driven around these protostars.

% TMC-1A and Uncertainties associated with mask size, P.A.
TMC-1A shows a spatial correlation of the velocity deviation with a steeper slope than the other two protostars, i.e., the velocity field around TMC-1A has a relatively larger gradient across a large spatial length compared to perturbations on small spatial scales. However, this could be due to systemic motions around the protostar, such as rotational and infalling motions, as the measured break radius would not correspond to the transition between the two regimes seen in IRAS 15398$-$3359 and L1527 IRS. Indeed, as seen in Figure \ref{fig:vprof_all}(c), the radial profile of the peak velocity shows a large velocity difference between inner 400--700 au radius and outer thousands au radius. Gas motions with high velocity around the protostar would cause a relatively large velocity gradient, and thus make the slope $\gamma_\mathrm{2D}$ steeper. On the other hand, the slope $\gamma_\mathrm{2D}$ is not expected to change significantly depending on the mask size for the turbulent velocity field. Thus, we calculated SFs with larger mask size, $r_\mathrm{mask}=$1700 au and 2400 au, to test whether the steeper slope found in TMC-1A is due to systemic motions around the protostar. The fitting results are summarized in Table \ref{tab:param_delvfit}. The obtained $\gamma_\mathrm{2D}$ with the largest mask size of 2400 au is much smaller than that derived with a mask of 800 au. We also measured $\gamma_\mathrm{2D}$ for IRAS 15398$-$3359 and L1527 IRS with the largest mask size of 2400 au, and obtained 0.5 and 0.6, respectively, which are comparable to the values derived with the smaller masks within fitting errors. These results, thus, suggest that the steep slope of 0.93 obtained with a mask size of 800 au for TMC-1A is due to systemic motions around the protostar that have relatively high velocity. The obtained $\gamma_\mathrm{2D}$ for TMC-1A with the mask size of 2400 au is consistent with the values measured for the other two sources and values reported for molecular clouds in previous works. However, the velocity structure could still originate from envelope components, as mentioned above, and it is difficult to conclude what motion is dominant for the velocity structure on $\sim$1000--10,000 au scales around TMC-1A.

% --------- Table ----------
% vpeak fit
\begin{table*}[thbp]
\begin{threeparttable}
    \centering
    \caption{Results of the fitting to the one-dimensional cuts of the moment 1 maps and the spatial correlation of the velocity deviation.}
    \label{tab:param_delvfit_sublin}
    \begin{tabular*}{2\columnwidth}{@{\extracolsep{\fill}}lccccc}
    \hline
    \hline
     Source & $r_\mathrm{mask}$ & $G$\tnote{1} & $v_0$\tnote{1} & $q$\tnote{2} & $\gamma_\mathrm{2D}$\tnote{2}  \\
      & (au)& ($\kmps~\mathrm{pc}^{-1}$) & ($\kmps$) & &   \\
    \hline
    IRAS 15398$-$3359 & 1200 & 0.5$\pm$0.2 & 5.203$\pm$0.005 & $-1.69\pm0.08$ & $0.5\pm0.3$ \\
    L1527 IRS & 1700 & 1.2$\pm$0.1 & 5.951$\pm$0.004 & $-1.59\pm0.04$ & $0.5\pm0.2$ \\
    \hline
    \end{tabular*}
    \begin{tablenotes}
    \item[1] $v(x)=G x + v_0$
	\item[2] $\delta v(\tau)=10^{q} (\tau/\mathrm{1000~\au})^{\gamma_\mathrm{2D}}$
	\end{tablenotes}
\end{threeparttable}
\end{table*}
% ---------------------

While the derived slopes for IRAS 15398$-$3359 and L1527 IRS are consistent with turbulent velocity field, we note that these results do not necessarily rule out the presence of any rotational motion on these scales. It would be possible that the velocity field of a hybrid of rigid-body rotation and turbulence exhibits a spatial correlation of the velocity deviation with a slope similar to the observed slopes, as presented in Figure \ref{fig:delv_diffmodels} in Appendix \ref{sec:app_tbmodel}. Hence, in order to assess the degree of possible rotational motions, a linear function is fitted to the one-dimensional velocity profile and SFs are calculated after subtracting the linear velocity gradients. Linear velocity gradients provide reasonable estimates on the projected rotational velocity, even in cases that the velocity field consists of rotational and turbulent motions \citep[][]{Stewart:2022aa}. The results of the fitting to the one-dimensional velocity profiles and spatial correlations of the velocity deviation are summarized in Table \ref{tab:param_delvfit_sublin}. The slope becomes slightly shallower compared to the slope without subtraction of the linear velocity gradients, as the overall velocity gradient responsible for the velocity deviation on a large scale is suppressed. The slopes are still consistent with turbulent velocity field for the driven turbulence case. We calculated the square of velocity deviation caused by the rotation ($\delta v_\mathrm{rot}^2$) and turbulence ($\delta v_\mathrm{turb}^2$) at a radius of 4000 au from the fitting results. For IRAS 15398$-$3359 and L1527 IRS, $\delta v_\mathrm{turb}^2$ is calculated to be $2 \times 10^{-3}$ km$^2$ s$^{-2}$ and $3 \times 10^{-3}$ km$^2$ s$^{-2}$ and $\sim$5--17 times larger than $\delta v_\mathrm{rot}^2$ of $10^{-4}$ km$^2$ s$^{-2}$ and $5 \times 10^{-4}$ km$^2$ s$^{-2}$, respectively. Although the rotational velocity is not corrected by the inclination angle, it affects $\delta v_\mathrm{rot}^2$ by only $0.8$--10\%, assuming the core rotational axis is the same as those of the outflow and disk, as the inclination angles of the outflows and disks in IRAS 15398$-$3359 and L1527 IRS are estimated to be $
\sim$70$^\circ$ and $
\sim$85$^\circ$, respectively \citep[$0^{\circ}$ for a pole-on configuration][]{Tobin:2008aa,Oya:2014aa, Aso:2017ab}. Thus, turbulent motions would be more dominant than rotational motions, even when the dense cores also possess a rotational motion.

In the two sources, IRAS 15398$-$3359 and L1527 IRS, the direction of the velocity gradient changes greatly depending on the spatial scale, as seen in Section \ref{subsec:ana_vgrad}. This trend is also consistent with turbulent velocity fields, where the directions of the specific angular momentum axis are expected to vary as a function of radius \citep{Joos:2013aa, Matsumoto:2017aa}, as is discussed in the next subsection in more detail. A recent simulation work by \cite{Misugi:2019aa} suggests that velocity fluctuation inside the filament, described with one-dimensional Kolmogorov power spectra, explains the $J/M$-$R$ relation measured in dense cores \citep[e.g.,][]{Goodman:1993aa, Tatematsu:2016aa}. Our results agree with this scenario. \cite{Hacar:2013aa} measured turbulent velocity of filaments to be typically sonic on 0.5 pc scale. Assuming Kolmogorov scaling law, it is expected that turbulent velocity is about 0.06 $\kmps$ at 4000 au, which is comparable to the values observed in IRAS 15398$-$3359 and L1527 IRS.

We note that, in another possible explanation of the velocity structures outside the break radius, \cite{Arroyo-Chavez:2022aa} recently demonstrated with numerical simulations that the $J/M$-$R$ relation observed in dense cores can be explained as a consequence of gravitational contraction with angular momentum loss via turbulent viscous. This scenario may be also applied to the relation of $j\propto r^{\sim1.6}$ found in the radial profile of the specific angular momentum measured in individual protostellar cores. However, as they calculated the angular momentum and the dense core radius in three dimension, and placed focus on the spatial scale of $\sim$0.1--1 pc, it is difficult to make a direct comparison between their simulations and our observed maps on $\sim$100-10,000 au scales. Simulated, projected maps on $\sim$1000-10,000 au scales will allow us to investigate how the velocity structures in individual dense cores look like in this scenario as well as whether they are consistent with our observational results.

% Subsection: comparision between sources
\subsection{Comparison with Other Sources}
\subsubsection{Break Radius}
The two Class 0 sources in our sample, L1527 IRS and IRAS 15398$-$3359, show a break at a radius of $\sim$1200 and 1700 au in the radial profile of the peak velocity, respectively. These breaks may correspond to the transition between the $j$-constant and $j$-increase regimes discussed in previous works \citep{Gaudel:2020aa, Sai:2022aa}, although the velocity structures outside the break radius around the two sources are not likely explained as rotational motion. \cite{Gaudel:2020aa} reported a break radius of 1600 au, which is similar to those of IRAS 15398$-$3359 and L1527 IRS, based on an averaged radial profile of the specific angular momentum of 12 Class 0 sources. \cite{Sai:2022aa} reported that the Class I protostar L1489 IRS exhibits a break at a radius of $\sim$2900 au, which is about two times larger than in L1527 IRS and IRAS 15398$-$3359. \cite{Pineda:2019aa} measured radial profiles of the specific angular momentum distributions within a radius of 800--10,000 au around two Class 0 sources (HH 211 and IRAS 03282$+$3035) and one first hydrostatic core candidate (L1451-mm), obtaining $j\propto r^{1.8}$ without a break. \cite{Lee:2010aa} found that the Class I protostar HH 111 exhibits rotation motion with a constant specific angular momentum within a radius of 2000--7000 au, suggesting that the transition between the two regimes occurs at a radius $\gtrsim$7000 au.

These comparisons of the break radius, including lower and upper limits and summarized in Table \ref{tab:rbreak}, suggest that the break radius between the two regimes is typically less than 2000 au for Class 0 sources. It is also found that the break radii of Class I sources are larger than those of Class 0 sources, although there are only two samples of Class I sources. This possible trend suggests that the extent of the infalling envelope with a constant specific angular momentum increases as the protostar evolves. This finding is consistent with results of an analytic core-collapse model calculated by \cite{Takahashi:2016aa}. They reported that the infalling gas element is extended in radial direction during the core collapse when the angular momentum is conserved, as the infalling velocity is faster at inner radii, which results in the flat radial distribution of the specific angular momentum \citep[see also][]{Yen:2011aa, Yen:2017aa}. These calculations predict that the extent of the flat radial distribution of the specific angular momentum increases with evolution, as larger specific angular momentum is brought in from larger radii at later evolutionary stages. It should be noted, however, that these models assume a very simple initial condition, such as a dense core rotating like a rigid-body without any turbulence and magnetic field. Numerical simulations with more realistic condition are required to further investigate whether the above picture is valid in turbulent dense cores.

\subsubsection{Velocity Structure Outside Break Radius}
The velocity structures outside the break radius around IRAS 15398$-$3359 and L1527 IRS are consistent with turbulence motion, as discussed in Section \ref{subsec:vfield_tb}. \cite{Gaudel:2020aa} also revealed that 12 Class 0 protostars exhibit large dispersions of the directions of velocity gradients over radii of 100--5000 au, and proposed the hypothesis of turbulent motions as the best interpretation of the velocity structures at scales of $>1600$ au. On the other hand, the velocity structure outside the break radius around L1489 IRS is interpreted as rotational motion by \cite{Sai:2022aa}, as the velocity gradient appears in the same direction over different spatial scales regardless of inside or outside of the break radius.

\begin{figure}[htbp]
\centering
\includegraphics[width=\columnwidth]{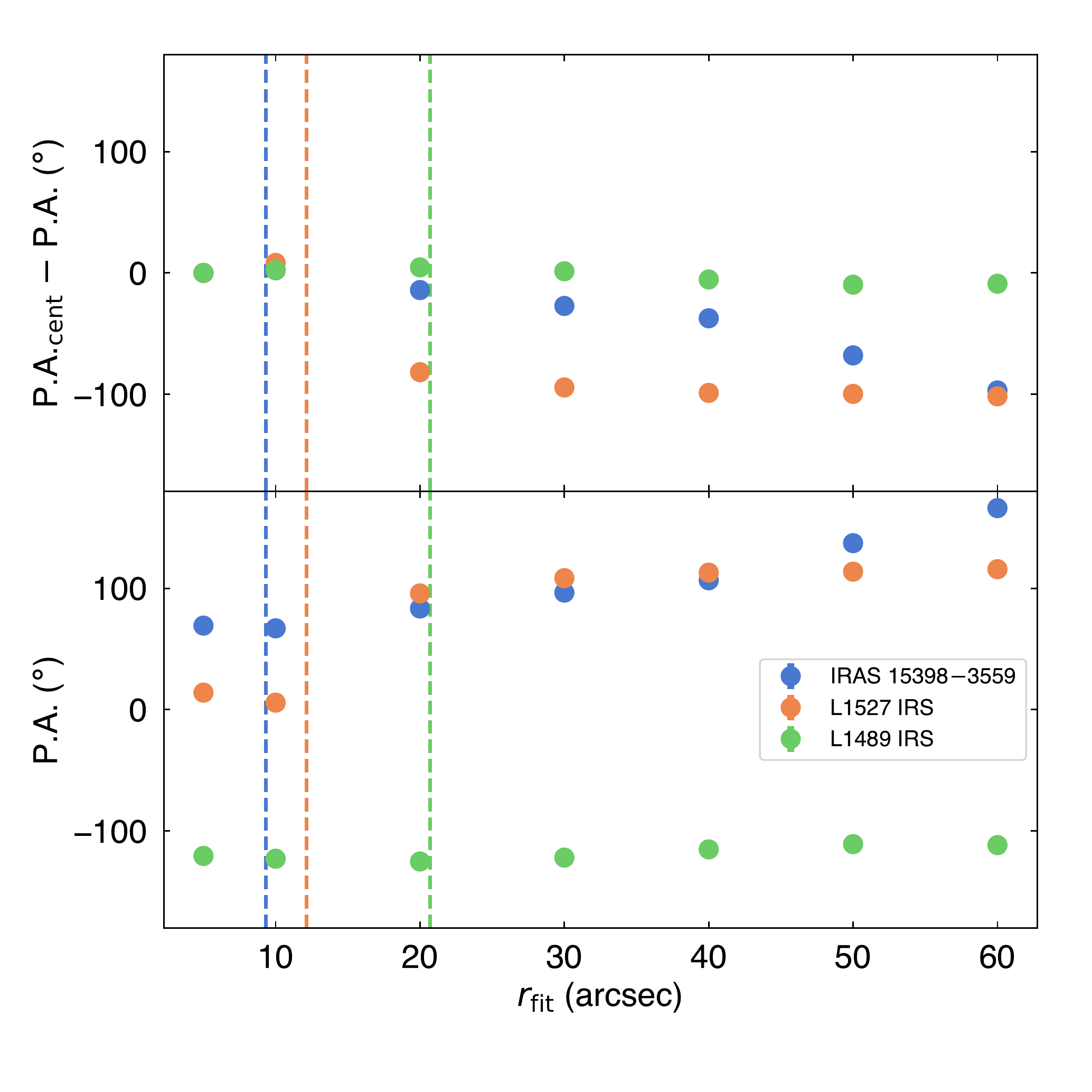}
\caption{Directions of velocity gradients measured within aperture areas with a radius of $r_\mathrm{fit}$. In the top panel, P.A.$_\mathrm{cent}$ is P.A. of the velocity gradient measured within $r_\mathrm{fit}$=$\ang[angle-symbol-over-decimal]{;;5}$. Vertical dashed lines denote the break radius measured in each source.}
\label{fig:vgrad_all}
\end{figure}

To compare the nature of velocity structures among different sources, the directions of the velocity gradients are measured as a function of the spatial scale, as shown in Figure \ref{fig:vgrad_all}, based on the measurements presented in Section \ref{subsec:ana_vgrad}. We exclude TMC-1A from comparison here, as we could not measure the break radius which likely corresponds to a transition between the $j$-constant and $j$-increase regimes in the source. For L1489 IRS, data points at $r_\mathrm{fit}$ of $\ang[angle-symbol-over-decimal]{;;20}$--$\ang[angle-symbol-over-decimal]{;;60}$ are compiled from measurements in \cite{Sai:2022aa}, where the magnitude and direction of the velocity gradient of L1489 IRS were measured with the same method used in Section \ref{subsec:ana_vgrad}. For data points of L1489 IRS at $r_\mathrm{fit} \leq \ang[angle-symbol-over-decimal]{;;10}$, we performed the same fitting as used in Section \ref{subsec:ana_vgrad} to the small-scale map of L1489 IRS, which is presented in Appendix \ref{sec:app_l1489}.

As seen in Figure \ref{fig:vgrad_all}, the directions of the velocity gradients of IRAS 15398$-$3359 and L1527 IRS significantly vary by about 90--100$^\circ$ from the smallest to largest scales. The dispersion of the directions is 34$^\circ$ for IRAS 15398$-$3359 and 45$^\circ$ for L1527 IRS. On the contrary, the dispersion of the directions of the velocity gradients of L1489 IRS is only $\sim\pm$5$^\circ$ across its break radius of 2900 au ($\sim\ang[angle-symbol-over-decimal]{;;21}$).

\begin{figure}[htbp]
\centering
\includegraphics[width=\columnwidth]{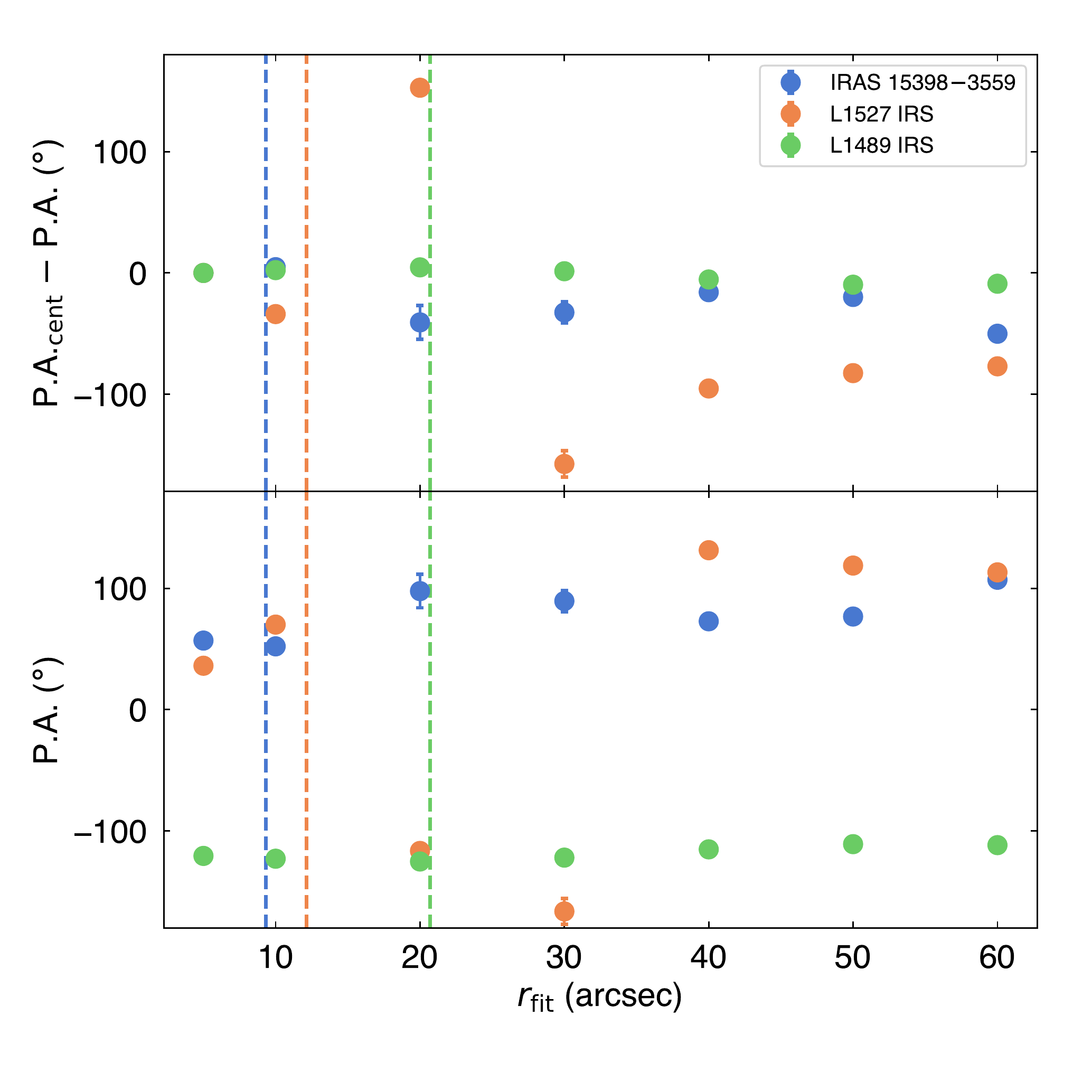}
\caption{Same as Figure \ref{fig:vgrad_all} but with measurements where outflows are masked for IRAS 15398$-$3359 and L1527 IRS.}
\label{fig:vgrad_wmask}
\end{figure}

One might wonder whether the large dispersion of the directions of the velocity gradients in IRAS 15398$-$3359 and L1527 IRS is because outflow components are dominant at particular spatial scales. Indeed, the position angles of the velocity gradient measured in IRAS 15398$-$3359 with the small-scale map are $\sim$70$^\circ$ and similar to that of the outflow \citep{Yen:2017aa}. The position angles of the velocity gradient in the large-scale map of L1527 IRS are $\sim$110$^\circ$ and are also close to the position angle of the outflow \citep[$\sim$90$^\circ$;][]{Tobin:2008aa}. To measure the orientations of the velocity gradients in the two sources minimizing the contribution of the outflows, we performed the same fitting to maps where area within $\pm$60$^\circ$ from the outflow axis is masked. The position angle of the outflow is assumed to be 50$^\circ$ for IRAS 15398$-$3359 and 90$^\circ$ for L1527 IRS, based on previous observations \citep{Tobin:2008aa, Yen:2017aa}. The fitting results with the masks are presented in Figure \ref{fig:vgrad_wmask}. The two sources still show a larger dispersion of the directions of the velocity gradients than L1489 IRS, even when emission associated with the outflows are not considered in the analysis. The dispersion measured in IRAS 15398$-$3359 with the outflow masking is 19$^\circ$, which is smaller than when measured without the outflow masking but still larger than in L1489 IRS. The direction of the velocity gradient varies by 50$^\circ$ at maximum. An inversion of the direction by 180$^\circ$ is seen at 20$''$ scale in L1527 IRS, as discussed in Section \ref{subsec:discussion_vpeakprof}. This is because an overall velocity gradient caused by the outflow is less significant. This leads to a larger dispersion of the orientation of 110$^\circ$ compared to when measured without the outflow masking.

\cite{Gaudel:2020aa} presented similar measurements in their Figure 19, showing that directions of velocity gradients vary more than 50$^\circ$ from 100 to 5000 au in all 12 Class 0 sources. Simulations of turbulent dense cores suggest that the directions of the specific angular momentum axis tends to vary as a function of radius \citep{Joos:2013aa, Matsumoto:2017aa}, as would be the case for IRAS 15398$-$3359 and L1527 IRS. These imply that the small dispersion of the direction of the velocity gradient outside the break radius around L1489 IRS, which is interpreted as rotational motion, is different from those around the other sources.

A possible explanation for the difference in the velocity structures outside the break radius between L1489 IRS and the other sources is the difference in the initial condition. Two out of three protostellar or prestellar sources, where \cite{Pineda:2019aa} have measured $j\propto r^{1.8}$, exhibit clear velocity gradients in directions perpendicular to their outflows. The velocity structure on several thousands au scales around the isolated Class 0 protostar B335 is well explained by the rigid-body rotation of the dense core \citep{Saito:1999aa, Kurono:2013aa}. These observations suggest that some of Class 0 sources has an ordered velocity structure with less variation in direction of local velocity gradients on several thousands au scales, which would be typically outside the infalling envelope with a constant specific angular momentum in the Class 0 phase. On the other hand, all 12 sources studied by \cite{Gaudel:2020aa} show a dispersion of the directions of velocity gradients at scales $>1600$ au. \cite{Tobin:2011aa} investigated the gas kinematics around 18 protostars on 1000 au to 0.1 pc scales, reporting that velocity gradients over $\sim$0.1 pc scales are often not normal to their outflows. These works with a large sample number suggest that the ordered velocity structure described by rotational motion is rare on $\sim$1000--10,000 au scales in the Class 0 phase. Furthermore, a warped disk structure is suggested around L1489 IRS, which is expected to form with a turbulent initial dense core \citep[][]{Sai:2020aa}. The other possibility explaining the difference in the velocity structures is the evolutionary effect. Rotational motion could be more dominant even outside the break radius at later evolutionary stages since energy injection by outflows are expected to decrease with time at the Class I phase by orders of magnitude \citep{Machida:2013aa} and turbulence would decay. If the difference of velocity structures outside the break radius is due to the evolutionary effect, Class I sources will tend to show more coherent velocity structures than Class 0 sources. Currently L1489 IRS is the only a sample of Class I protostars where the break radius is measured. Comparison of the velocity structures between Class 0 and I sources with more samples is needed to investigate the evolution of the velocity structures in more detail.

% ------- figure -------

% ----------------------

% ----------- Table: results ------------
% Velocity gradients
\begin{table*}[htbp]
\begin{threeparttable}
    \centering
    \caption{Break radii estimated in protostellar and prestellar sources.}
    \label{tab:rbreak}
    \begin{tabular*}{2\columnwidth}{@{\extracolsep{\fill}}lccc}
    \hline
    \hline
     Evolutionary stage & Source/number of sources & $\rbreak$ (au) & Reference \\
    \hline
    FHSC\tnote{1} candidate/Class 0 & 3 sources & $<$1000 & \cite{Pineda:2019aa} \\
    Class 0 & 12 sources & 1600$\pm$300 & \cite{Gaudel:2020aa} \\
    Class 0 & IRAS 15398$-$3359 & 1180$\pm90$ & This work \\
    Class 0 & L1527 IRS & 1700$\pm50$ & This work \\
    Class I & L1489 IRS & 2900$\pm$200 & \cite{Sai:2022aa} \\
    Class I & HH 111 & $>$7000 & \cite{Lee:2010aa}\\
    \hline
    \end{tabular*}
    \begin{tablenotes}
        \item[1]First Hydrostatic Core
    \end{tablenotes}
\end{threeparttable}
\end{table*}
% ---

\section{Summary and Conclusion \label{sec:summary}}
We have observed the three protostars, IRAS 15398$-$3359, L1527 IRS and TMC-1A, with the ALMA 12-m array, ACA 7-m array, APEX and the IRAM 30m telescope in the C$^{18}$O $J=$2--1 emission. The kinematic structures on $\sim$100--10,000 au scales have been investigated with the two different types of combined maps: small-scale maps covering regions within a radius of $\sim$800 au around the protostars with spatial resolutions of $\sim$110 au, and large-scale maps covering wider areas within a radius of $\sim$9000 au with spatial resolutions of $\sim$1000 au. Our main results are summarized as follows:

\begin{enumerate}
    \item In the small-scale maps, two out of the three sources, L1527 IRS and TMC-1A, exhibit clear velocity gradients that likely originate from rotational motions of the disks and envelopes. IRAS 15398$-$3359 does not show a clear velocity gradient in the direction perpendicular to the primary outflow, but shows a velocity gradient along the primary outflow direction. In the large-scale maps, on the other hand, velocity gradients are less clear and velocity structures are more perturbed in all the sources.
    \item To investigate the radius at which the dominant motion changes from rotational motion of the envelope to others, the peak velocity $v_\mathrm{peak} = |\vlsr - \vsys|$ was measured as a function of radius from $\sim$400 to 9000 au using PV diagrams cut along the disk major axis. IRAS 15398$-$3359 and L1527 IRS exhibit a break at a radius of $\sim$1200 and 1700 au, respectively, in the radial profiles of the peak velocity. Fitting of a double-power law function results in the power-law index of $-1.7\pm{0.3}$ and $-1.38\pm 0.03$ inside the break radius, which are smaller than the power-law index of $-1$ expected for the infalling envelope under conservation of specific angular momentum. These slopes would be interpreted as indicating rotational motions of the envelope with contamination of gas motion on larger spatial scales along the line of sight. The power-law indices outside the break radius are, on the other hand, $\sim$0.46--0.68, which are similar to the slope of the $J/M$-$R$ relation of dense cores ($v\sim J/M/R \propto R^{0.6}$). TMC-1A shows a break in its radial profile of the averaged peak velocity at a radius of $\sim$800 au, and the power-law indices of $\sim-4.1$ and $\sim-0.30$ inside and outside the break radius, respectively. These slopes are consistent with neither the slopes expected for the infalling envelope conserving specific angular momentum nor the $J/M$-$R$ relation of dense cores.
    \item In order to probe the origin of the velocity structures outside the break radius, spatial correlations of the velocity deviation were investigated through the second-order structure function (SF) of the centroid velocity. The spatial correlations of the velocity deviation, $\delta v \propto \tau^{\sim0.6}$, were found in IRAS 15398$-$3359 and L1527 IRS. The values of the derived slopes are consistent with predictions of turbulence models within an uncertainty associated with the projection effect, assuming that turbulence is being driven around the protostars. These results may suggest that the velocity structures outside the break radius around the two sources originate from turbulence. TMC-1A exhibits a steeper slope, 0.93, which is likely due to contamination of systemic motions around the protostar.
    \item The break radius estimated for IRAS 15398$-$3359 and L1527 IRS is similar to those derived in previous works. A comparison of the break radii between Class 0 and I sources suggests a possible evolutionary trend that the extent of the infalling envelope with a constant specific angular momentum increases with time, although the number of Class I samples is limited. The velocity structure around the Class I source L1489 IRS appears more coherent across the break radius compared to those around other Class 0 sources. This difference would be due to the difference of the initial condition or turbulence decay at later evolutionary stages. Further observations of more samples at different evolutionary stages are needed to distinguish these two possibilities.
\end{enumerate}

This paper makes use of the following ALMA data: ADS/JAO.ALMA \#2013.1.00879.S, 2013.1.01086.S, and 2019.1.01063.S. ALMA is a partnership of ESO (representing its member states), NSF (USA), and NINS (Japan), together with NRC (Canada), MOST and ASIAA (Taiwan) and KASI (Republic of Korea), in cooperation with the Republic of Chile. The Joint ALMA Observatory is operated by ESO, AUI/NRAO, and NAOJ. This work is based on observations carried out under project number 044-14 and 129-19 with the IRAM 30-m telescope. IRAM is supported by INSU/ CNRS (France), MPG (Germany), and IGN (Spain). This publication is based on data acquired with the Atacama Pathfinder Experiment (APEX). APEX is a collaboration between the Max-Planck-Institut fur Radioastronomie, the European Southern Observatory, and the Onsala Space Observatory. N.O. is supported by National Science and Technology Council (NSTC) in Taiwan through the grant NSTC 109-2112-M-001-051 and 110-2112-M-001-031. H.-W.Y. acknowledges support from Ministry of Science and Technology (MOST) in Taiwan through the grant MOST 110-2628-M-001-003-MY3 and from the Academia Sinica Career Development Award (AS-CDA-111-M03). A.M. is supported by the European Research Council (ERC Starting Grant MagneticYSOs with grant agreement no. 679937).

\facility{ALMA, IRAM-30m, APEX}

\software{CASA \citep{McMullin:2007aa}, GILDAS (\url{http://www.iram.fr/IRAMFR/GILDAS}), MIRIAD \citep{Sault:1995aa}, Numpy \citep{Oliphant:2006aa,van-der-Walt:2011aa}, Scipy \citep{Virtanen:2020aa}, Astropy \citep{Astropy-Collaboration:2013aa,Astropy-Collaboration:2018aa}, Matplotlib \citep{Hunter:2007aa}}

\restartappendixnumbering
\appendix

\section{Velocity Channel Maps \label{subsec:app_channelmaps}}
\begin{figure*}[htbp]
\centering
\includegraphics[width=2\columnwidth]{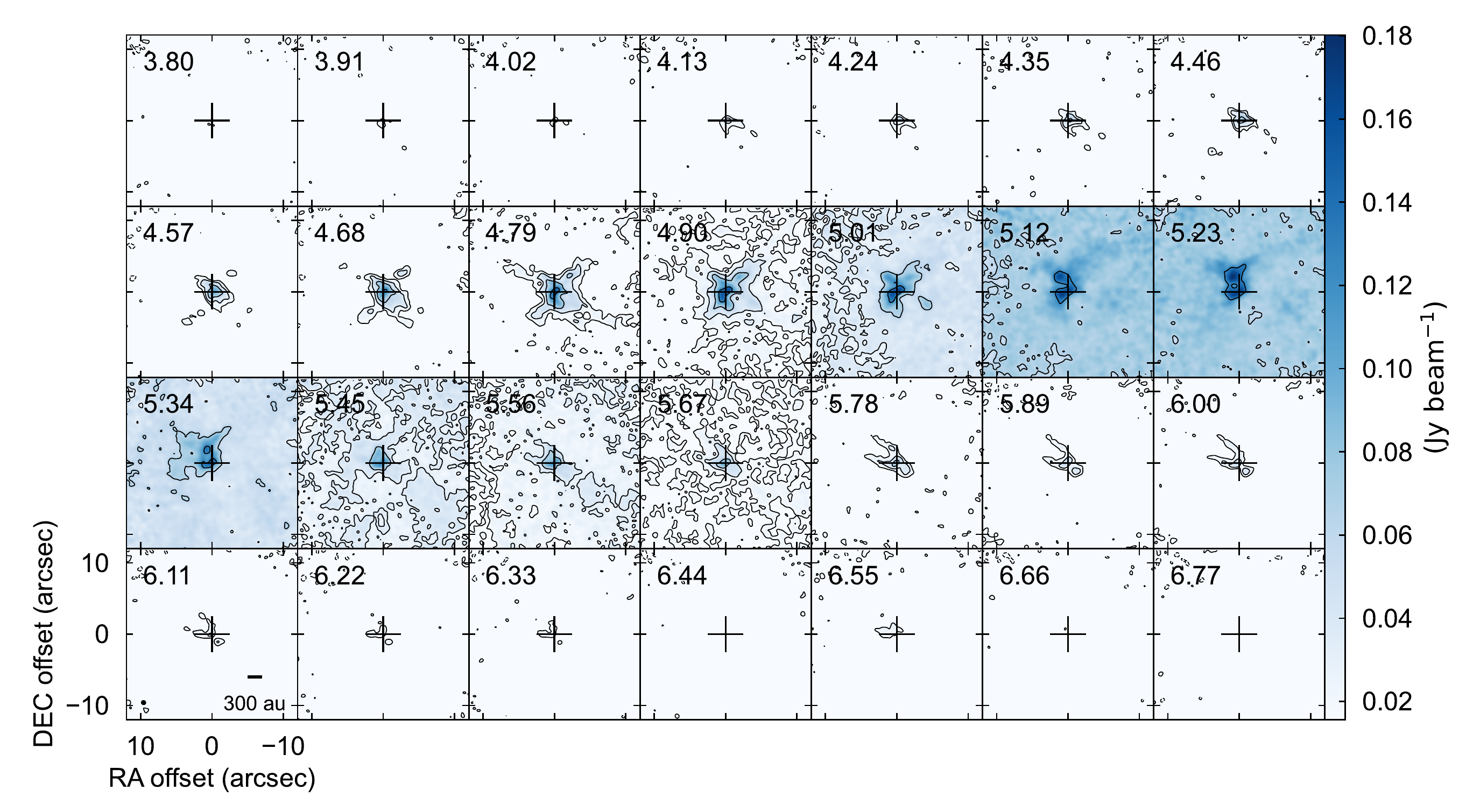}
\caption{Small-scale channel maps of the C$^{18}$O $J=2$--1 emission of IRAS 15398$-$3359. Contour levels are 3, 6, 12, 24, ... $\times \sigma$, where 1$\sigma=5.2~\mjypbm$. The labels in the top-left corner indicate the LSR velocity of each channel in $\kmps$. Crosses at the center and a filled ellipse in the bottom-left corner denote the protostellar position and the beam size, respectively. The systemic velocity of IRAS 15398$-$3359 is 5.18 $\kmps$.}
\label{fig:channel_iras15398}
\end{figure*}

\begin{figure*}[htbp]
\centering
\includegraphics[width=2\columnwidth]{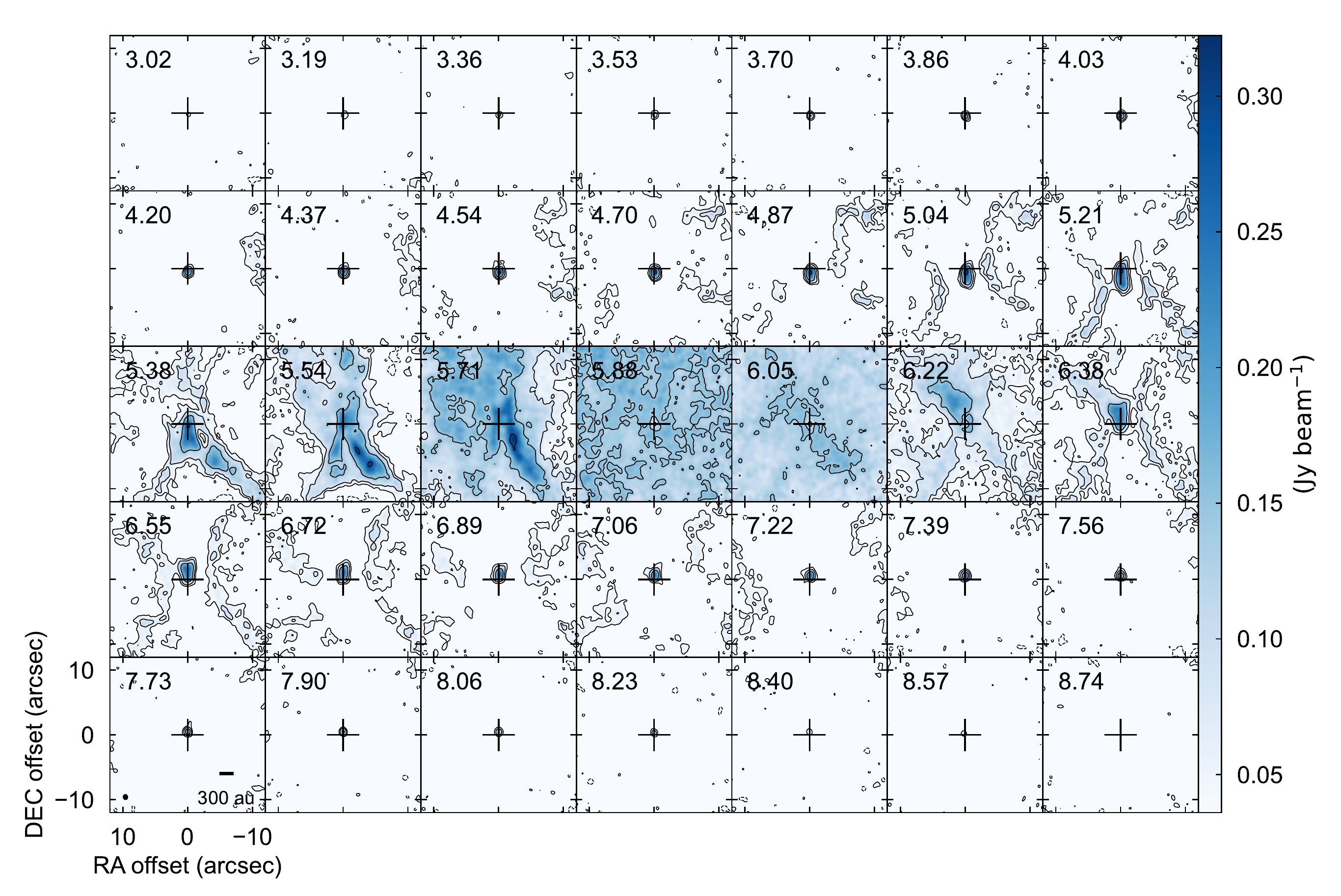}
\caption{Same as Figure \ref{fig:channel_iras15398} but for L1527 IRS with 1$\sigma=12~\mjypbm$. Maps are shown in steps of two channels. The systemic velocity of L1527 IRS is 5.8 $\kmps$.}
\label{fig:channel_l1527}
\end{figure*}

\begin{figure*}[htbp]
\centering
\includegraphics[width=2\columnwidth]{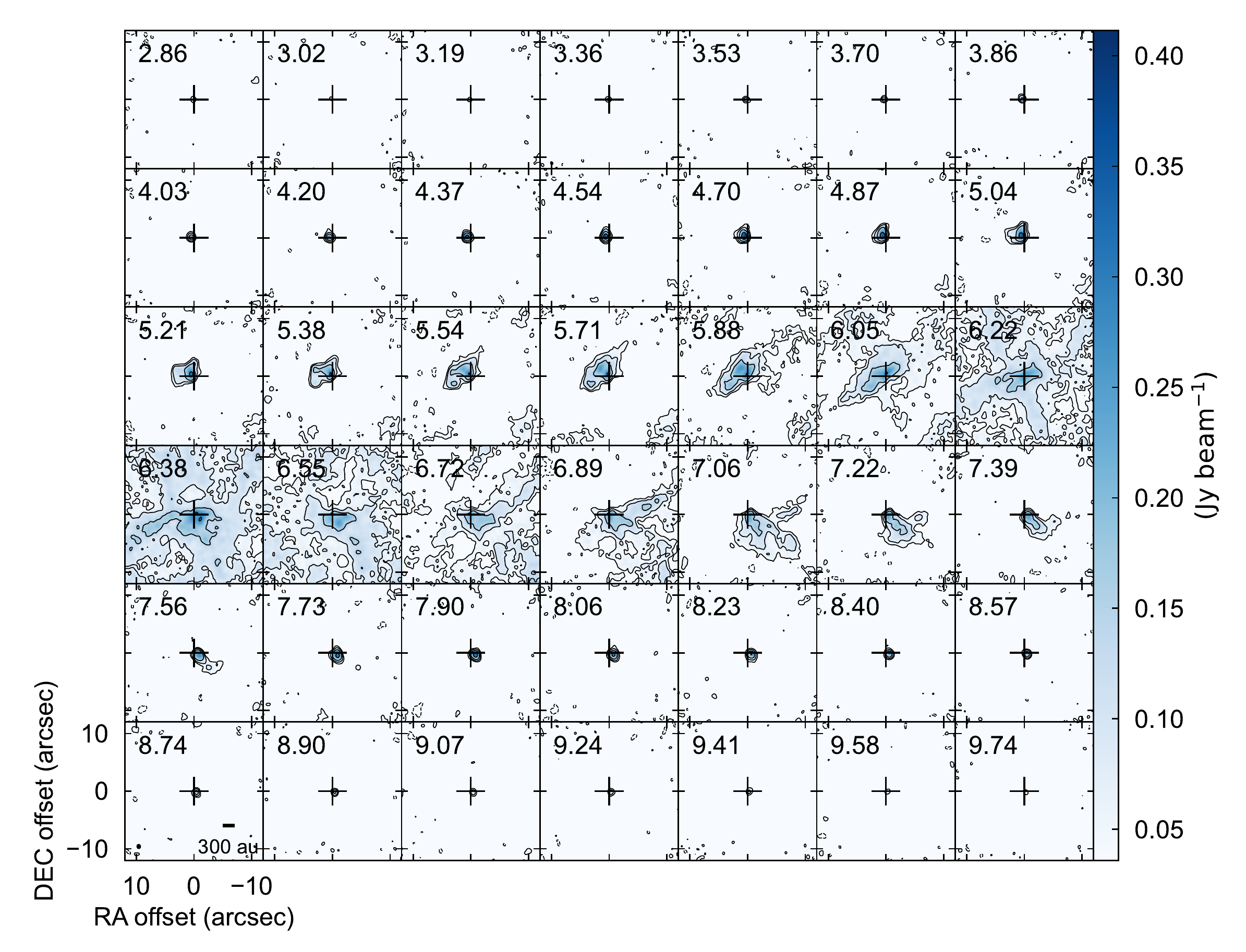}
\caption{Same as Figure \ref{fig:channel_iras15398} but for TMC-1A with 1$\sigma=12~\mjypbm$. Maps are shown in steps of two channels. The systemic velocity of TMC-1A is 6.4 $\kmps$.}
\label{fig:channel_tmc1a}
\end{figure*}

\begin{figure*}[htbp]
\centering
\includegraphics[width=2\columnwidth]{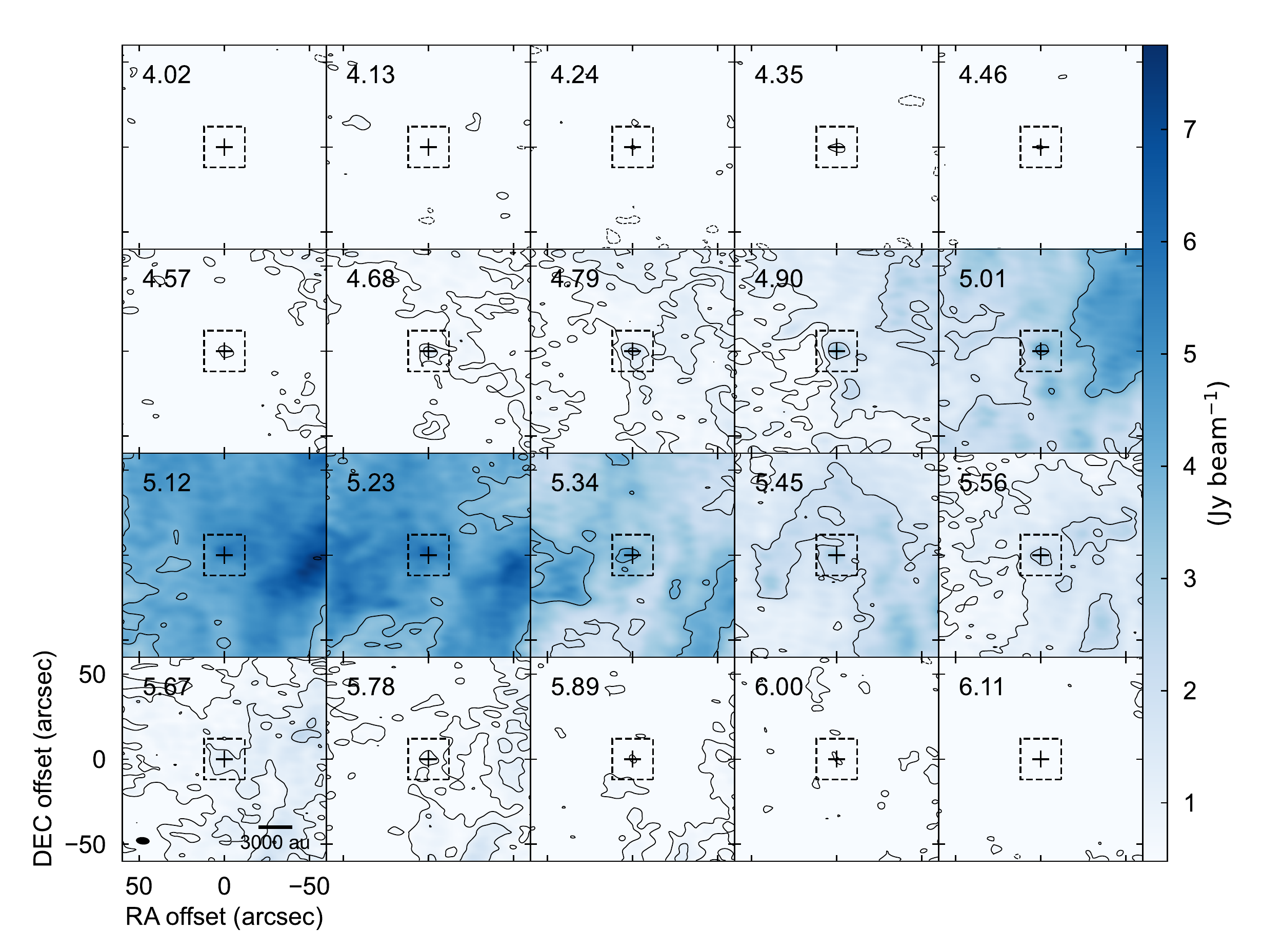}
\caption{Large-scale channel maps of the C$^{18}$O 2--1 emission of IRAS 15398$-$3359. Contour levels are 3, 6, 12, 24, ... $\times\sigma$, where 1$\sigma=0.16~\jypbm$. Dashed boxes show the size of the small-scale map ($\ang[angle-symbol-over-decimal]{;;12}$ in radius). The labels in the top-left corner indicate the LSR velocity of each channel in $\kmps$. Crosses at the center and a filled ellipse in the bottom-left corner denote the protostellar position and the beam size, respectively. The systemic velocity of IRAS 15398$-$3359 is 5.18 $\kmps$.}
\label{fig:channel_ls_iras15398}
\end{figure*}

\begin{figure*}[htbp]
\centering
\includegraphics[width=2\columnwidth]{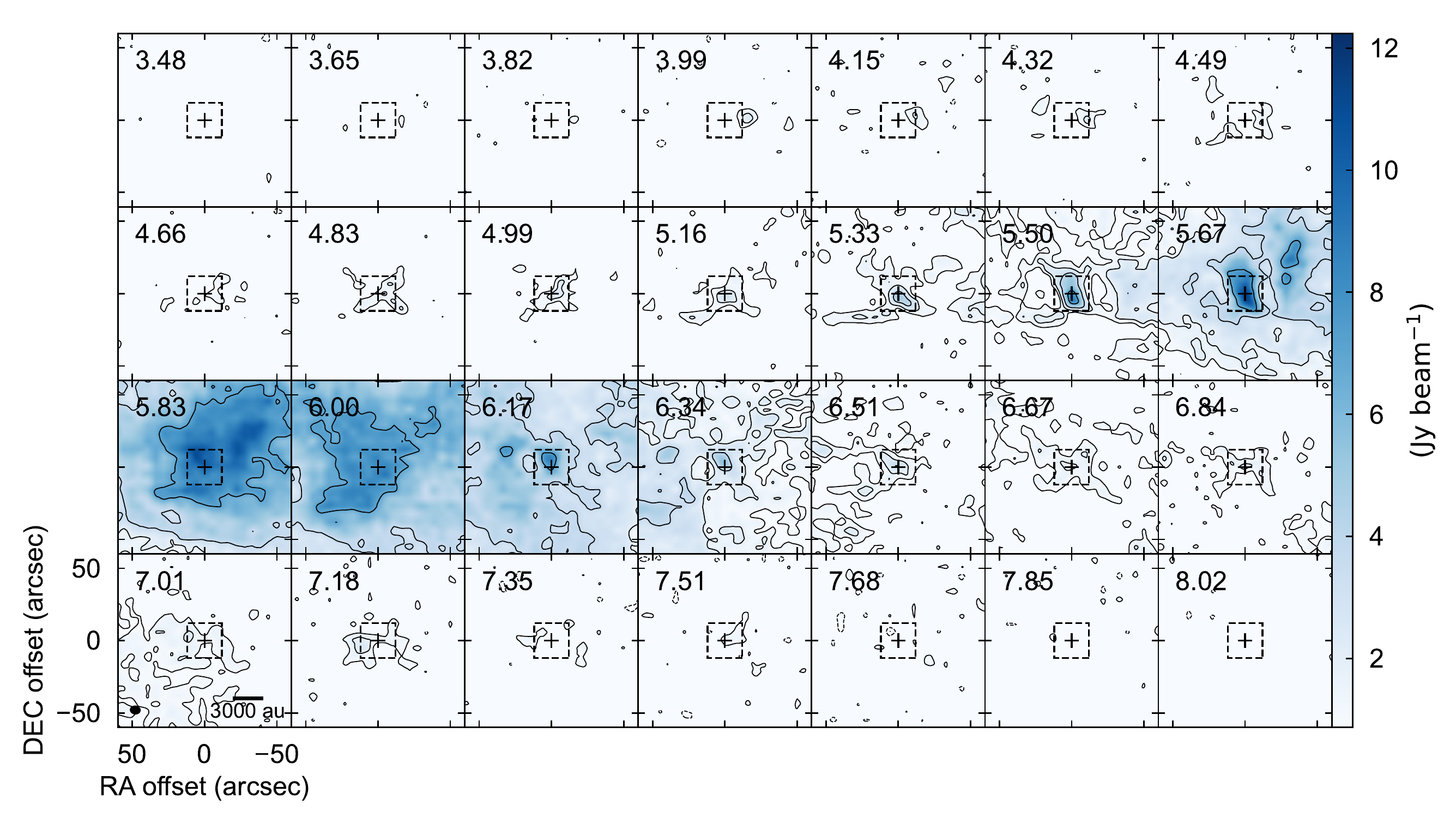}
\caption{Same as Figure \ref{fig:channel_ls_iras15398} but for L1527 IRS with 1$\sigma=0.29~\jypbm$. Maps are shown in steps of two channels. The systemic velocity of L1527 IRS is 5.8 $\kmps$.}
\label{fig:channel_ls_l1527}
\end{figure*}

\begin{figure*}[htbp]
\centering
\includegraphics[width=2\columnwidth]{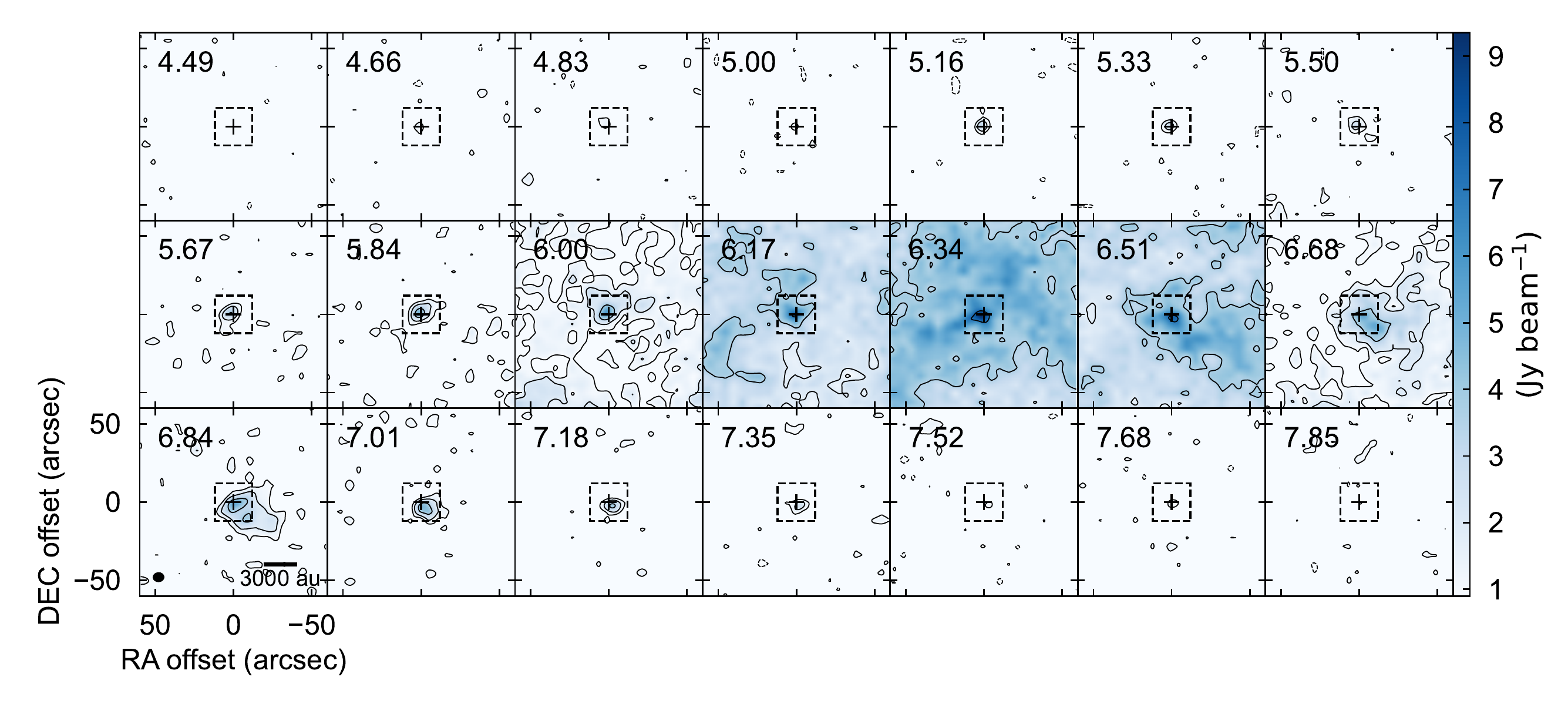}
\caption{Same as Figure \ref{fig:channel_ls_iras15398} but for TMC-1A with 1$\sigma=0.30~\jypbm$. Maps are shown in steps of two channels. The systemic velocity of TMC-1A is 6.4 $\kmps$.}
\label{fig:channel_ls_tmc1a}
\end{figure*}

Velocity channel maps of the C$^{18}$O $J=2$--1 emissions of IRAS 15398$-$3359, L1527 IRS and TMC-1A are presented in Figure \ref{fig:channel_iras15398} to Figure \ref{fig:channel_ls_tmc1a}.

% channel maps
% iras 15398

% l1527

% tmc-1a
%

% Large-scale channel maps

% l1527

% tmc-1a
% ---------------------------

\newpage

% turbulence model
\section{One-Dimensional Model of Turbulent Velocity Field   \label{sec:app_tbmodel}}
The slope of the spatial correlation of the velocity deviation obtained from one-dimensional velocity fields is examined using models of the one-dimensional turbulent velocity field in order to assess the robustness of the discussion in Section \ref{subsec:vfield_tb}.

The model turbulent velocity field is computed based on the method described by \cite{Dubinski:1995aa}. The Kolomogorov-like power spectrum of the turbulent velocity field is given by
\begin{align}
    P_v(k) = \langle |v(k)|^2 \rangle = C (k^2 + k_\mathrm{min}^2)^{p_k/2},
\end{align}
where $k$ is the spatial frequency, $k_\mathrm{min}$ is the spatial frequency of the largest scale of turbulence and $p_k$ is the slope of the power spectrum. Here, $\langle \rangle$ denotes the ensemble average. The given power spectrum results in the spatial correlation of the velocity deviation, $\delta v\propto \tau^{(p_k - n)/2}$, where $n$ is the dimension. The velocity field in $k$ space is calculated as follows \citep{Myers:1999aa}:
\begin{align}
    v(k) = v_k \exp{(ik x)} \exp{(i\phi_k}),
\end{align}
where $v_k$ is the amplitude determined by the power spectrum and $\phi_k$ is a random phase uniformly distributed from 0 to 2$\pi$. The velocity field in the real space is obtained as the inverse Fourier transform of $v(k)$.

\begin{figure*}[htbp]
\centering
\includegraphics[width=2\columnwidth]{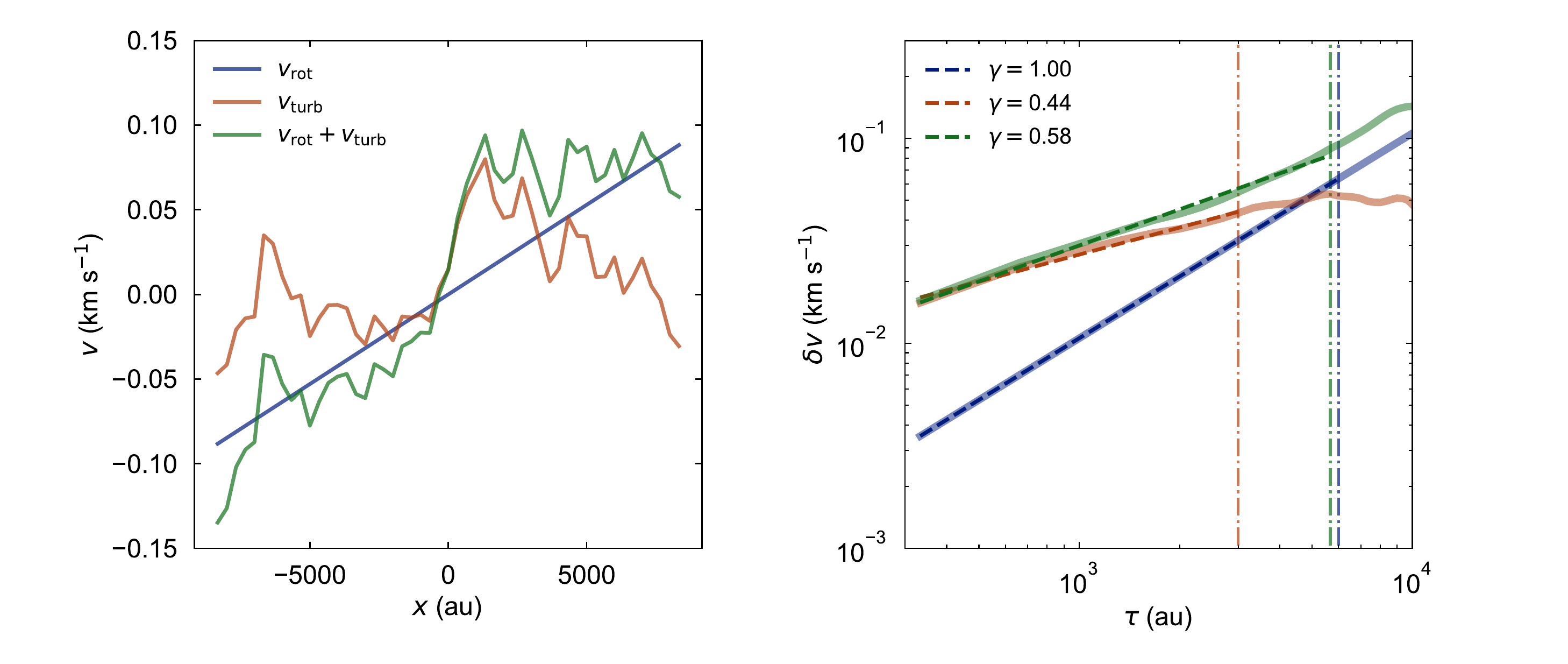}
\caption{(Left) Examples of one-dimensional velocity fields of a pure rotational motion, a pure turbulent motion, and hybrids of rotational and turbulent motions. (Right) Spatial correlations of the velocity deviation for the one-dimensional velocity fields shown in the left panel. Solid lines show the structure functions calculated from the velocity fields, the vertical dash-dotted lines show $\tau_0$ estimated from their ACFs, and dashed lines show the best-fit power-law relations. Labels at the top left corner indicate the best-fit slopes.}
\label{fig:delv_diffmodels}
\end{figure*}

In our calculations, we adopted $p_k=2$, corresponding to the Larson's law $\delta v\propto \tau^{0.5}$. To simulate the current observations, we have set the domain size and the grid number to $L=17,000$ au and 51, respectively, which are comparable to those of the sampled one-dimensional slice of the moment 1 maps. The largest scale of turbulence, $\lambda_\mathrm{max}= 1/k_\mathrm{min}$, is taken to be $2L$. The constant $C$ in the power spectrum is chosen so that the velocity deviation over the entire length is $\sim$0.03 $\kmps$, which is comparable to that of the observations. An example of the turbulent velocity field and the correlation between the spatial scale and velocity deviation is shown in Figure \ref{fig:delv_diffmodels}.

We also present the one-dimensional velocity field and the spatial correlation of the velocity deviation for cases of a rigid-body rotation and a hybrid of the turbulence and rigid-body rotation in Figure \ref{fig:delv_diffmodels}. The model calculations demonstrate that the slope of the spatial correlation of the velocity deviation for the rigid-body rotation is one and much steeper than that for the turbulent velocity field. A hybrid velocity field of the rigid-body rotation and turbulence, on the other hand, can show a slope similar to that for the turbulent velocity field.

\begin{figure}[htbp]
\centering
\includegraphics[width=\columnwidth]{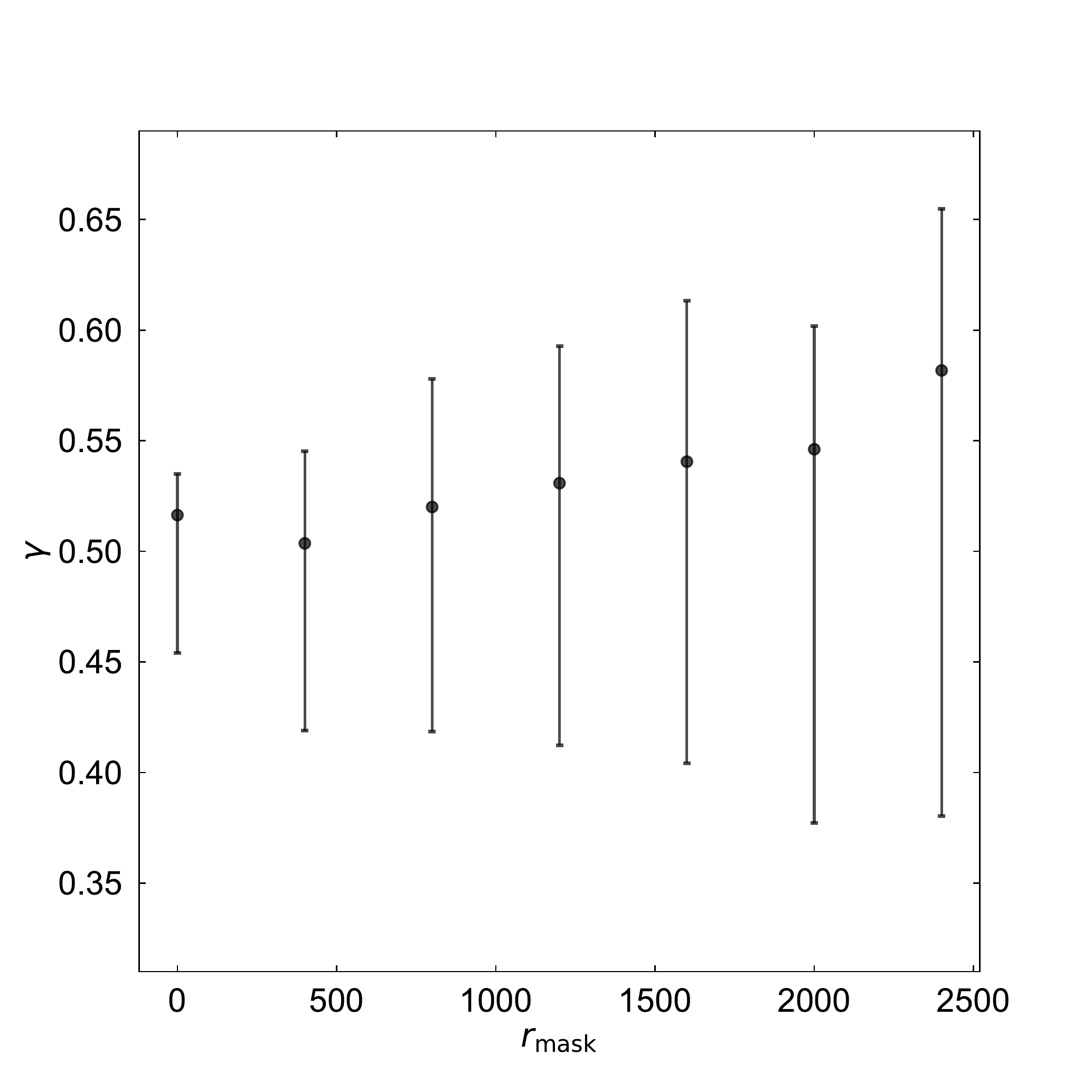}
\caption{Comparison between the mask size and obtained slopes of the spatial correlations of the velocity deviation for the one-dimensional turbulent velocity field.}
\label{fig:delv_tbmodel}
\end{figure}

To assess the effect of masking, we applied the same analysis as that performed in Section \ref{subsec:vfield_tb} to the model velocity field with and without a mask. The uncertainty of the measured slope is evaluated with the Monte-Carlo (MC) method by performing the analysis with 1000 model velocity fields, which have random phases, and obtained as 68\% highest density interval of the posterior probability distribution. Figure \ref{fig:delv_tbmodel} shows the relation between the mask size and obtained slopes of the spatial correlation of the velocity deviation. A larger mask size increases the uncertainty of the slope but the slope is always around the expected value of 0.5, which indicates that the masking does not significantly change the obtained slope. According to these calculations, the uncertainty of the slope associated with the mask size for IRAS 1539$-$3359 and L1527 IRS is about 0.1, assuming that their true slope is around 0.5.

\newpage
\section{The small-scale map of L1489 IRS \label{sec:app_l1489}}
\begin{figure}[htbp]
\centering
\includegraphics[width=\columnwidth]{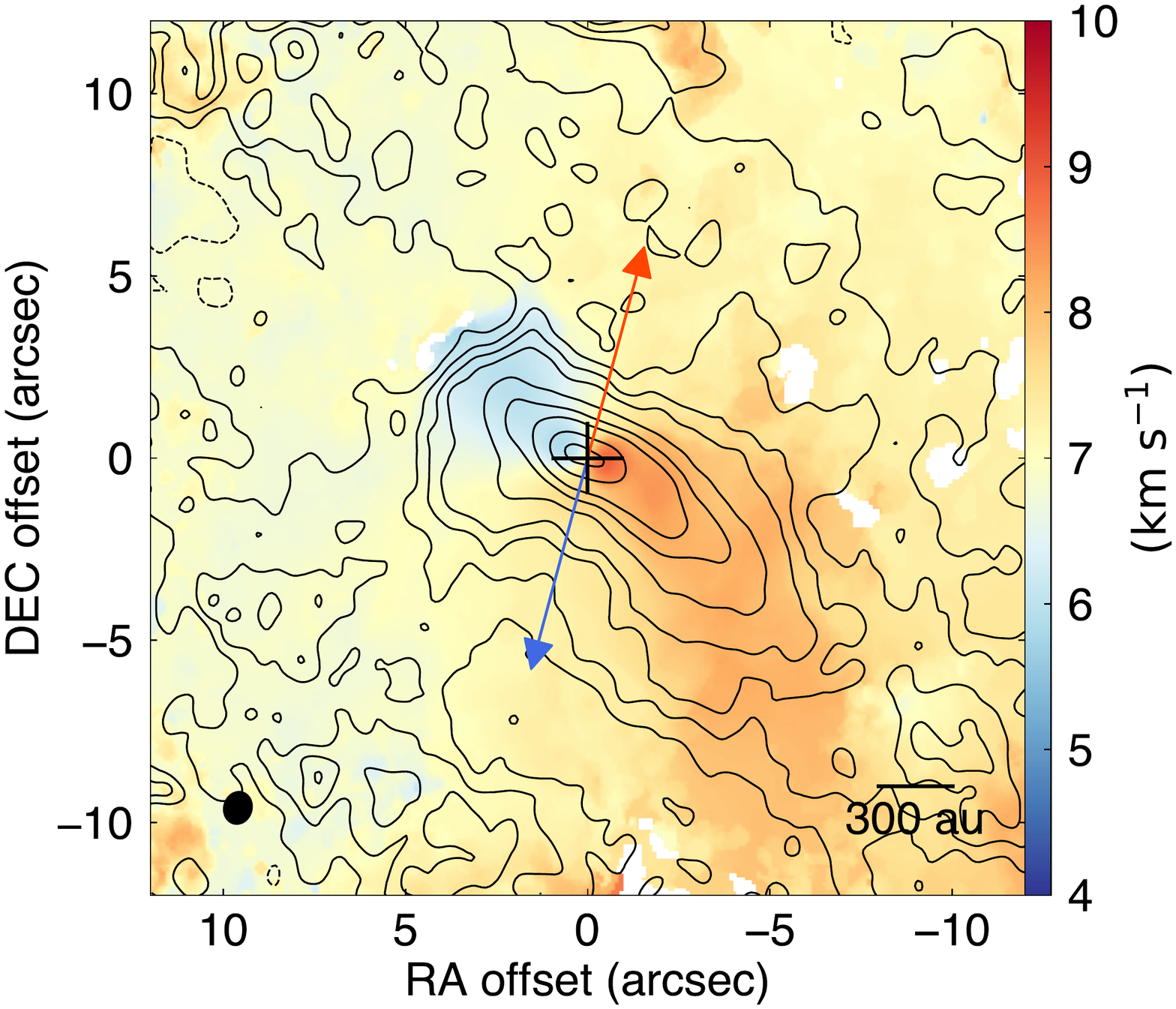}
\caption{Same as Figure \ref{fig:momentmaps} but for the small-scale map of L1489 IRS.}
\label{fig:app_ssmap_l1489}
\end{figure}

For comparison of velocity gradients, we produced the small-scale map of the Class I protostar L1489 IRS combining the observational data of ALMA 12-m array, ACA 7-m array and IRAM-30m, whose observational details are presented in \cite{Sai:2020aa} and \cite{Sai:2022aa}. The imaging and data combining were performed with the same parameters and method described in Section \ref{subsec:obs_combmaps}. The velocity resolution and rms of the resultant map are 0.084 $\kmps$ and 8.1 $\mjypbm$, respectively. The synthesized beam size is $\ang[angle-symbol-over-decimal]{;;0.94} \times \ang[angle-symbol-over-decimal]{;;0.81}$ ($-16^\circ$). The integrated intensity and centroid velocity maps are presented in Figure \ref{fig:app_ssmap_l1489}.

We measured the magnitude and direction of the velocity gradient in the small-scale map of L1489 IRS with the method described in Section \ref{subsec:ana_vgrad}. The fitting result is summarized in Table \ref{tab:app_vgrad_l1489}.

% ------- Figure --------
% L1489 IRS
% --------------

% ------- Table ------
% L1489 IRS
\begin{table}[htbp]
\centering
\begin{threeparttable}
    %\scalefont{0.9}
    \centering
    \caption{Result of the two-dimensional linear fitting to the small-scale centroid velocity map of L1489 IRS.}
    \label{tab:app_vgrad_l1489}
    \begin{tabular*}{0.8\columnwidth}{@{\extracolsep{\fill}} lccc}
    \hline
    \hline
    $r_\mathrm{fit}$ & $v_0$\tnote{a} & $G$ & $\theta$ \\
     ($''$) & ($\mathrm{km~s^{-1}}$) & ($\mathrm{km~s^{-1}~pc^{-1}}$) & ($^\circ$) \\
    \hline
    5 & 7.20 & $283.8 \pm 0.2$ & $-120.61 \pm 0.04$ \\
    10 & 7.17 & $77.13 \pm 0.04$ & $-122.94 \pm 0.03$ \\
    \hline
    \end{tabular*}
    \begin{tablenotes}
	\item[a] Fitting errors of $v_0$ are less than 0.1 \%.
	\end{tablenotes}
\end{threeparttable}
\end{table}
% --------------------

\bibliographystyle{aasjournal}
\bibliography{reference}



\end{document}